\documentclass[a4paper,12pt]{article}

\usepackage[hmargin=.7in,vmargin=1.1in]{geometry}
\usepackage{indentfirst}
\usepackage{amsfonts}
\usepackage{mathrsfs}
\usepackage{amsmath}
\usepackage{amssymb}
\usepackage{authblk}
\usepackage{cite}
\usepackage{xcolor}
\usepackage{mathtools}
\usepackage{tensor}
\usepackage{physics}
\usepackage{graphicx}
\usepackage{bm}
\usepackage{tikz}
\usepackage{upgreek}
\usepackage{braket}
\usepackage{color,soul}
\usepackage{csquotes}
\usepackage{caption}
\usepackage{subcaption}
\usepackage{tabularx}
\usepackage{simplewick}
\usepackage[section]{placeins}
\usepackage{enumerate}
\usepackage{comment}
\newcommand{\pp}{\uppi}

\newcommand\mi{\mathrm{i}}
\newcommand\me{\mathrm{e}}

\newcommand{\dif}{\mathrm{d}}
\newcommand{\red}[1]{\textcolor{red}{#1}}
\DeclareMathOperator{\diag}{diag}

\usepackage[bookmarksnumbered=true,bookmarksopen=true]{hyperref}
 \hypersetup{colorlinks,
             linkcolor=[rgb]{0,0.3,0.6}, 
             citecolor=[rgb]{0,0.3,0.6}, 
             urlcolor=[rgb]{0,0.3,0.6}}

\title{\Large Low-finesse scattering and non-stationary dispersive dynamics of gravitational wave echoes}

    \author[1,2]{Han-Wen Hu\thanks{huhanwen@itp.ac.cn}}
    \author[1,2]{Cheng-Jun Fang\thanks{fangchengjun@itp.ac.cn}}
    \author[1,2,3]{Zong-Kuan Guo\thanks{guozk@itp.ac.cn}}

\affil[1]{\normalsize{\em Institute of Theoretical Physics, Chinese Academy of Sciences, P.O. Box 2735, Beijing 100190, China}}	

\affil[2]{\normalsize{\em School of Physical Sciences, University of Chinese Academy of Sciences, No.19A Yuquan Road, Beijing 100049, China}}	

\affil[3]{\normalsize{\em School of Fundamental Physics and Mathematical Sciences, Hangzhou Institute for Advanced Study, University of Chinese Academy of Sciences, Hangzhou 310024, China}}

\numberwithin{equation}{section}

\date{}

\begin{document}

\maketitle

\begin{abstract}
We study environmental echoes induced by a weak potential barrier outside a Schwarzschild black hole. 
In the low-finesse limit, the time domain response is governed by a sequence of transient wave packets formed by finite round-trip scattering, rather than steady state cavity modes. 
We establish quantitative criteria for the breakdown of the steady state resonance picture, dictated by frequency domain spectral aliasing and time domain truncation from the black hole power law tail. 
Based on non-stationary dispersive dynamics, we analytically derive the arrival time gliding, central frequency drift, and dispersion driven asymmetric tails of these echoes. 
Accordingly, we construct a five-parameter analytical reconstruction whose overlap with the finite difference time domain extracted first echo is nearly identical to that obtained from the full transfer-function reconstruction. 
Our results demonstrate that early low-finesse environmental echoes must be theoretically modeled as non-stationary transient scattering signals.
\end{abstract}

\section{Introduction}

Gravitational wave (GW) detections have confirmed that remnant black holes (BHs) consistently align with the theoretical predictions of Kerr geometry \cite{LIGOScientific:2016lio,LIGOScientific:2019fpa,LIGOScientific:2021sio,KAGRA:2021vkt,LIGOScientific:2025wao,LIGOScientific:2025rid}.
However, to resolve quantum gravity challenges like the information loss paradox \cite{Hawking:1976ra,Mathur:2009hf,Almheiri:2020cfm}, various models—including fuzzballs \cite{Mathur:2005zp,Mathur:2008nj,Mathur:2009hf}, firewalls\cite{Almheiri:2012rt,Ori:2012jx}, gravastars\cite{Mazur:2001fv,Mazur:2004fk}, traversable wormholes\cite{Damour:2007ap,Bueno:2017hyj,Churilova:2021tgn,Yang:2024prm}—propose replacing the classical horizon with a reflective boundary \cite{Cardoso:2016rao,Cardoso:2019rvt}.
This boundary, paired with the intrinsic potential barrier of BH, forms a resonant cavity that traps gravitational radiation to produce near-horizon echoes \cite{Cardoso:2016oxy,Cardoso:2016rao,Mark:2017dnq,Maselli:2017tfq,Zimmerman:2023hua}, whose waveforms in rotating backgrounds are further modulated by greybody factors, spin, and complex reflectivity \cite{Maggio:2019zyv}.
A closely related classical cavity-trapping mechanism appears in ultracompact stars, where an interior potential well, bounded by the exterior curvature barrier and by the regular center or stellar surface, can temporarily confine spacetime perturbations and generate trapped $w$-modes or resonant/interference peaks in particle-scattering spectra \cite{Kokkotas:1995av,Tominaga:1999iy,Ferrari:2000sr}.
Beyond quantum corrections, realistic astrophysical BHs are often embedded in macroscopic environments like dark matter halos, boson clouds, or accretion disks \cite{Isi:2018pzk,Zhu:2020tht,Baumann:2022pkl,Bertone:2024rxe,Kazempour:2024lcx,Tomaselli:2024dbw,Yuan:2025fde,Berti:2025hly}. 
These environments can introduce an additional weak potential barrier significantly lower than the intrinsic barrier of BH, driving the system into a low-reflectivity limit and generating late-time environmental echoes \cite{Naidoo:2021rzw,Becar:2023zbl,Liu:2023vno,Pezzella:2024tkf,Daghigh:2025wcw}.
In analogy to classical optics, we define this low-reflectivity regime as the low-finesse limit. 
Under this limit, extreme single-trip energy dissipation prevents the formation of coherent standing waves. Consequently, the system degenerates from a steady state resonant cavity into a finite-round-trip scattering process, where trapped radiation manifests as a sequence of transient wave packets rather than discrete cavity eigenmodes.

The contrast between high-finesse resonance and low-finesse scattering highlights a fundamental contradiction between the mathematical poles of the frequency domain Green’s function and their physical observability. 
While pseudospectrum analysis and spectral instability reveal that BH quasinormal modes (QNMs) can undergo pole migration under microscopic environmental perturbations\cite{Jaramillo:2020tuu,Jaramillo:2021tmt,Destounis:2021lum,Cheung:2021bol,Courty:2023rxk,Destounis:2023ruj,Cai:2025irl,Hu:2025efp}, the observable time domain response is not dictated solely by these asymptotic poles.
Instead, time domain evolution is strictly constrained by the causal structure, finite propagation time, and the explicit excitation efficiency of each individual mode \cite{Hui:2019aox,Daghigh:2020jyk,Kyutoku:2022gbr,Yang:2024vor,Cardoso:2024mrw,Oshita:2025ibu}.
In an external cavity with low reflectivity, the echo sequence undergoes rapid decay and is quickly submerged by the late-time power law tail of the prompt ringdown \cite{Qian:2020cnz,Qian:2024zvq,Kyutoku:2022gbr,Carullo:2023tff}. 
Consequently, the steady state resonance picture—which relies on infinite round-trip scattering—is physically inadequate for describing echoes in macroscopic low-finesse environments \cite{Berti:2022xfj,Cardoso:2024mrw}.

Consequently, an inherent structural mismatch exists between current echo search paradigms and the genuine dynamics of low-finesse environmental echoes. 
Existing methods typically rely on either explicit near-horizon waveform templates or model-independent searches for uniform comb-like spectra and long-lived QNMs \cite{Wang:2018gin,Lo:2018sep,Uchikata:2019frs,Maggio:2019zyv,Ren:2021xbe,Wu:2023wfv}. 
Both approaches fundamentally presume a steady state structure characterized by equally spaced pulses, uniform frequency combs, or resolvable narrow resonances.
However, the intrinsic potential barrier near the photon sphere acts as a strongly frequency-dependent greybody filter. 
High-frequency components preferentially penetrate the horizon during round-trip reflections, inevitably causing a systematic redshift in the surviving signal's dominant frequency band \cite{Nakano:2017fvh,Maggio:2019zyv,Conklin:2019fcs,Okabayashi:2024qbz,Rosato:2025byu}.
In the low-finesse limit, this spectral drift ceases to be a minor perturbation, it drives the echo sequence into a non-stationary regime characterized by strong dispersion and a finite number of pulses. 
Thus, early environmental echoes must be theoretically characterized as transient scattering events rather than steady state resonances.

Building on the transfer matrix and Dyson series formalisms, we model early echoes as successive transient scattering events rather than superpositions of steady state QNM poles \cite{Mark:2017dnq,Correia:2018apm}. Accordingly, our core contributions are threefold:
First, we establish quantitative criteria for the breakdown of the steady state resonance picture, dictated by spectral aliasing and power law tail truncation. 
Second, we develop a non-stationary dispersive theory to analytically capture the arrival time gliding, central frequency drift, and dispersion driven Airy tails of transient echoes.
Third, we encapsulate this non-stationary dispersive structure into a five parameter analytical reconstruction. 
Compared against the full transfer-function benchmark, this parameterization captures the dominant non-stationary evolution in low-finesse environments, a regime where traditional rigid translation approximations fundamentally fail.

The paper is organized as follows: Sec.\ \ref{sec:II} presents the time domain phenomenology of low-finesse environmental echoes via numerical evolution. 
Sec.\ \ref{sec.III} establishes quantitative criteria for the breakdown of the steady state resonance picture. 
Sec.\ \ref{sec:IV} analytically models the non-stationary dispersive evolution of these transient echoes. 
Sec.\ \ref{sec:V} introduces a five-parameter template and verifies its validity against the exact transfer function. 
Sec.\ \ref{sec:VI} provides a brief conclusion and discussion.

\section{Phenomenology of low-finesse environmental echoes}\label{sec:II}

This section establishes a phenomenological benchmark: 
in the low-finesse limit, environmental echoes manifest as transient wave packets formed by finite scattering events, exhibiting non-stationary evolution with increasing echo order.

\subsection{Transfer matrix theory of gravitational echos}

\begin{figure}[!htb]
    \centering
    \includegraphics[width=0.6\linewidth]{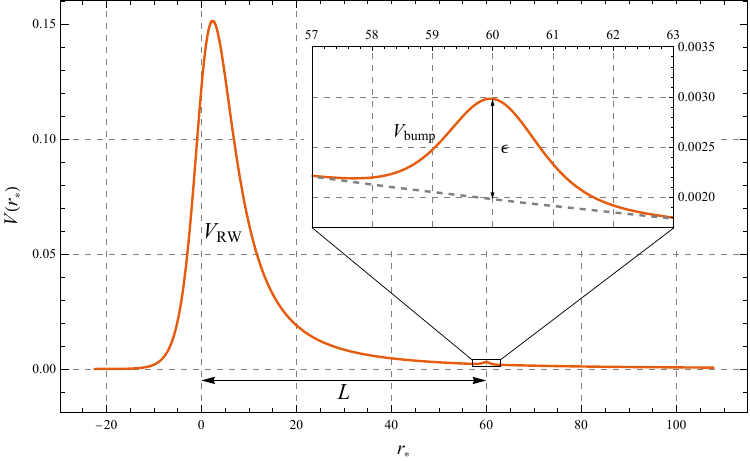}
    \caption{
    Schematic effective potential for the low-finesse environmental-echo model. 
    The near-horizon photon-sphere barrier is described by the Regge-Wheeler potential $V_{\rm RW}$, while the distant environmental structure is represented by the weak P\"oschl-Teller bump $V_{\rm bump}=\epsilon\sech^2(r_*-L)$.
    \red{Here we adopt typical parameters $\epsilon=10^{-3}$ and $L=60$ for the quadrupole mode $\ell=2$, where the total effective potential $V_{\rm RW} + V_{\rm bump}$ corresponds to the axial potential $V_{\rm ax}$ given in Eq.\ \eqref{eq:axial-potential}}.
    }
    \label{fig:rw-bump-potential}
\end{figure}

To characterize this mechanism, we employ the one-dimensional scattering picture of BH perturbation theory. 
In the Schwarzschild background, axial gravitational perturbations in the frequency domain are described by a master variable $\Psi(r_*, \omega)$, which satisfies the Regge-Wheeler equation
\begin{equation}\label{eq:RW-eq}
    \frac{\dif^2 \Psi(r_*, \omega)}{\dif r_*^2} + \left[\omega^2 - V_{\text{eff}}(r_*) \right] \Psi(r_*, \omega) = 0,
\end{equation}
where $r_*$ is the tortoise coordinate.
The total effective scattering potential $V_{\rm eff}(r_*) = V_{\rm RW}(r_*) + V_{\rm bump}(r_*)$ consists of the standard Regge-Wheeler barrier $V_{\rm RW}(r_*)$ and an external perturbation $V_{\rm bump}(r_*)$ induced by macroscopic environmental matter (e.g., dark matter halos). 
We adopt a P\"oschl-Teller soft barrier model for the environment
\begin{equation}
    V_{\rm bump}(r_*)=\epsilon \sech^2 (r_*-L),
\end{equation}
where $\epsilon$ parameterizes the barrier strength and $L$ denotes its central position, dictating the macroscopic round-trip time delay ($t \simeq 2L$). 
A schematic illustration of this macroscopic two-barrier structure is presented in Fig.\ \ref{fig:rw-bump-potential}.
To mimic realistic astrophysical scenarios in the extreme low-finesse limit, $\epsilon$ is assigned small values.

Here we address the astrophysical interpretation of the prescribed potential bump. 
As emphasized in Ref.~\cite{Cardoso:2024mrw}, a minor deformation in the Regge-Wheeler potential does not trivially equate to the presence of a small amount of matter. 
Introducing a secondary maximum far from the primary light-ring barrier imposes a new length scale and establishes a two-barrier cavity. 
If such a potential bump is reconstructed from a static spherical geometry, this additional extremum in the eikonal limit can correspond to a secondary light ring; 
consequently, the associated spacetime may deviate significantly from a perturbatively small deformation of the Schwarzschild background. 
This subtlety is explicitly manifest in the axial gravitational potential for matter-supported spherical backgrounds,
\begin{equation}\label{eq:axial-potential}
    V_{\rm ax} = f(r)\left[\frac{\ell(\ell+1)}{r^2} -\frac{6M}{r^3} + 4\pp \left(\rho-p_r\right) \right],
\end{equation}
where the deformation of the wave potential is dictated by the mass aspect and stresses, rather than the matter density alone. 
Therefore, the bump amplitude $\epsilon$ employed here should be interpreted as a phenomenological scattering-strength parameter. 
For instance, freezing the background to the Schwarzschild geometry and adopting a minimal anisotropic-fluid model with $p_r=\chi\rho$ yields $V_{\rm bump} \simeq 4\pp f(1-\chi)\rho$ \cite{Tian:2025plunging}. 
This implies a matter mass variation $\Delta M \sim (1-\chi)^{-1}\int \dif r_*\ r^2 V_{\rm bump} $, explicitly demonstrating that translating the amplitude $\epsilon$ into a physical matter mass is highly sensitive to the assumed stress tensor and the spatial location of the bump. 
For our P\"oschl-Teller bump with $L= 180 M$, one has $r_L \simeq 171.1M$; 
taking $\chi=1/2$ gives $\Delta M_\chi/M\simeq1.17(\epsilon/10^{-5})$, with an equivalent effective compactness of $C_{\rm eff} \equiv \Delta M_\chi/ r_L \simeq 6.8 \times 10^{-3} (\epsilon/ 10^{-5})$.
This illustrates that despite the non-negligible cumulative mass ($\sim 1.17M$), its dilution over an expansive spatial region allows that the effective compactness remains small. 
\red{
It should be noted that for $\epsilon \geq 10^{-4}$, the equivalent mass of the perturbation significantly exceeds that of the central black hole. 
We emphasize that these extended values are adopted exclusively for the mathematical demonstration of the spectral dynamics and multi-echo evolution, rather than representing realistic astrophysical configurations.
}
Given these complexities, in the remainder of this work, $\epsilon$ is utilized purely to parameterize the phenomenological strength of the added exterior scattering barrier, while its conversion to an astrophysical mass or surface density depends entirely on the specific matter model.

To solve this cavity scattering problem, we apply the transfer matrix method \cite{Ianniccari:2024ysv,Rosato:2025byu,Wu:2025sbq}. 
The asymptotic wave amplitudes are encapsulated in the state vector $\mathbf{\Phi}(r_*)$
\begin{equation}
    \mathbf{\Phi}(r_*) = \begin{pmatrix} A(r_*) \\ B(r_*) \end{pmatrix}
\end{equation}
with $\Psi(r_*) \sim A(r_*) \me^{\mi \omega r_*} + B(r_*) \me^{-\mi \omega r_*}$.
The global scattering properties of the system are fully encoded in the global transfer matrix $\mathbf{M}(\omega)$, which relates the scattering state at the horizon $\mathbf{\Phi}_{\text{H}}$ to the scattering state at spatial infinity $\mathbf{\Phi}_{\text{inf}}$,
\begin{equation}
    \mathbf{\Phi}_{\text{inf}} = \mathbf{M}(\omega) \cdot \mathbf{\Phi}_{\text{H}} = \left( \mathbf{T}_{\rm bump} \cdot \mathbf{P}_L \cdot \mathbf{T}_{\rm RW} \right) \cdot \mathbf{\Phi}_{\text{H}},
\end{equation}
where $\mathbf{T}$ is the local transfer matrix corresponding to each potential barrier, and $\mathbf{P}_{\rm L} = \diag(\me^{\mi \omega L}, \me^{-\mi \omega L})$ is the propagation matrix that accounts for the geometric phase accumulated by the wave packet propagating in the resonant cavity with length $L$.
Imposing the purely outgoing boundary condition at infinity and introducing the initial transient ringdown excitation $S_{\text{in}}(\omega)$, we obtain the global transfer function for the observed waveform $\tilde{\Psi}_{\text{obs}}(\omega)$
\begin{equation}\label{eq:trans-func}
    \tilde{\Psi}_{\text{obs}}(\omega) = \frac{T_{\rm bump}(\omega) R_{\rm RW}(\omega) \me^{\mi \omega L}}{1 - R_{\rm RW}(\omega) R_{\rm bump}(\omega) \me^{2 \mi \omega L}} S_{\text{in}}(\omega),
\end{equation}
where $R_X$ and $T_X$ denote the reflectivity and transmittance of the potential barrier $X$, respectively.
The complex poles of transfer function are defined by the condition $1 - \mathcal{L}(\omega_n) = 0$, where $\mathcal{L}(\omega)\equiv R_{\rm RW}(\omega) R_{\rm bump}(\omega) \me^{2 \mi \omega L}$ is the loop propagator. 
These poles correspond to the QNMs of the system, which describe the steady state resonance formed by GWs after infinite round-trip reflections.
However, for the early time domain response, the steady state resonance has not yet been established. In the low-reflectivity limit ($|R_{\rm bump}| \ll 1$), a more reasonable description is to expand the above transfer function into a geometric series
\begin{align}\label{eq:trans-func-series}
    \tilde{\Psi}_{\text{obs}}(\omega) & = \tilde{\Psi}_{\text{prompt}}(\omega) \sum_{n=0}^{\infty} \left[R_{\rm RW}(\omega) R_{\rm bump}(\omega) \me^{2 \mi \omega L} \right]^n \nonumber \\
    &\equiv \tilde{\Psi}_{\text{prompt}}(\omega) \sum_{n=0}^{\infty} \left[\mathcal{L}(\omega)\right]^n \equiv \sum_{n=0}^{\infty} \tilde{\Psi}_n,
\end{align}
where the term with $n=1$ corresponds to the first echo, with the explicit form $\tilde{\Psi}_{1} = \tilde{\Psi}_{\text{prompt}} \cdot \mathcal{L}(\omega)$.
This confirms that early echoes fundamentally manifest as transient wave packets strictly modulated by the loop propagator $\mathcal{L}(\omega)$, rather than resonant excitations of cavity eigenmodes.

\subsection{Time domain and frequency domain analysis}

We define ``viable echoes" in the time domain, which must simultaneously satisfy two phenomenological criteria:
\begin{enumerate}
    \item Sufficiently separated in time from the prompt ringdown signal;
    \item The main oscillation and envelope characteristics of the wave packet can be stably extracted within the corresponding time window $W_n$, and are not completely submerged by the late-time power law tail of the prompt ringdown.
\end{enumerate}
For each viable echo within $W_n$, we define the peak arrival time $t_n^{\rm peak}$ of the $n$-th wave packet as
\begin{equation}
t_n^{\rm peak} = \arg\max_{t \in W_n} |\Psi(t)|.
\end{equation}
From this, we extract the effective time interval between adjacent echoes $\Delta t_n = t_{n+1}^{\rm peak} - t_n^{\rm peak}$ and the peak amplitude $A_n = |\Psi(t_n^{\rm peak})|$, which parameterizes the wave packet's energy.

\begin{figure}[!htb]
    \centering
    \begin{subfigure}[t]{0.475\textwidth}
         \centering
         \includegraphics[width=\textwidth]{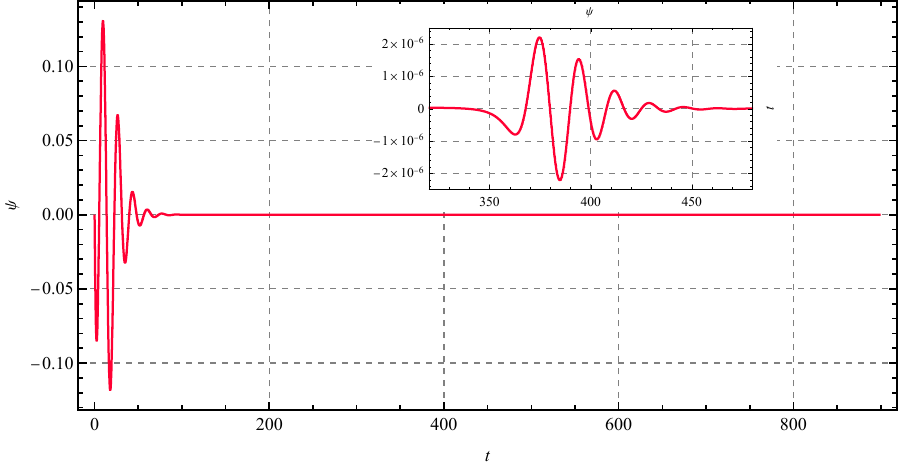}
         \caption{Linear time domain waveform.}
         \label{fig:1-echo-normal}
     \end{subfigure}
     \begin{subfigure}[t]{0.475\textwidth}
         \centering
         \includegraphics[width=\textwidth]{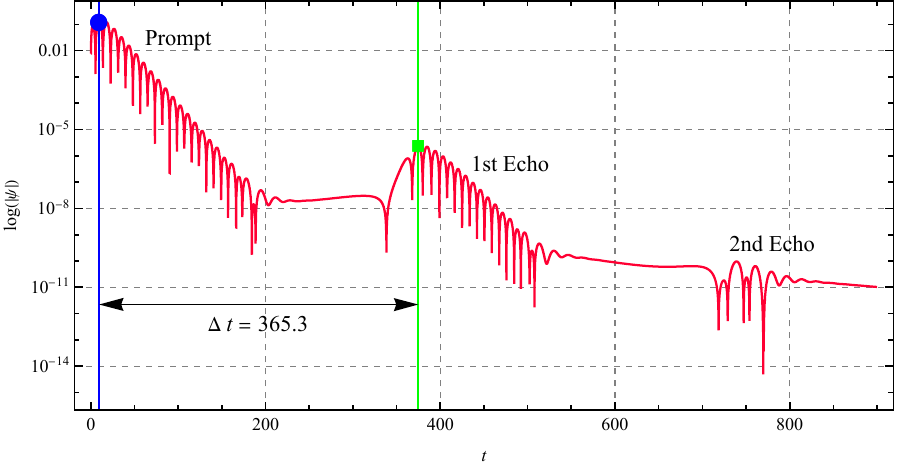}
         \caption{Logarithmic amplitude $|\Psi(t)|$, distinguishing the prompt ringdown and the isolated echoes.}
         \label{fig:1-echo-log}
     \end{subfigure}
    \begin{subfigure}[t]{0.475\textwidth}
         \centering
         \includegraphics[width=\textwidth]{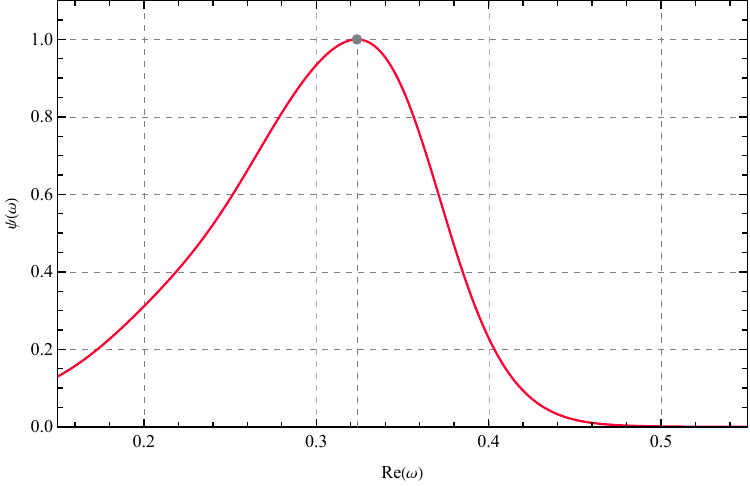}
         \caption{Fourier spectrum of the extracted first echo. The grey dashed line marks the peak frequency at $\omega \simeq 0.324$.
         Here we take the time window for the first echo as $t\in[360,720]$.}
         \label{fig:1-echo-FFT}
     \end{subfigure}
    \caption{Phenomenology of the environmental echo in the extreme low-finesse limit (with barrier parameters $L=180$ and $\epsilon=10^{-5}$). 
    Panels (a) and (b) illustrate the time domain separation of the first echo wave packet from the prompt ringdown and the power law tail. 
    Panel (c) highlights the smooth, broadband nature of its frequency spectrum, phenomenologically confirming the absence of steady state discrete resonances.}
    \label{fig:1-echo}
\end{figure}

We numerically integrate the Regge-Wheeler equation \eqref{eq:RW-eq} using a null-like finite difference time domain (FDTD) method. 
To satisfy the first viability criterion, we fix the cavity length at $L=180$ and extract all waveforms at a far-field observation point $r_*^{\rm obs}=300 \gg L$. 
The barrier strength is set to $\epsilon = 10^{-5}$ to firmly place the system in the low-finesse limit typical of realistic astrophysical environments.
The resulting time domain waveforms are shown in Fig.\ \ref{fig:1-echo-normal} and Fig.\ \ref{fig:1-echo-log}. 
The first echo emerges at $t \simeq 2L$ as a clearly isolated wave packet above the relatively flat power law tail, confirming it as an independent scattering event. 
However, accumulated high-order dispersion from multiple reflections severely distorts the morphology of the second echo, inducing highly asymmetric tails. 

To isolate individual echoes from contamination by the power-law tail and non-dispersive leading-edge spikes, while ensuring consistent results through a uniform reference wave packet width, we apply fixed-time hard truncation and extract a segment of length $360$ for each wave packet.
To compute the spectrum $\tilde{\Psi}_n(\omega)$ with high frequency resolution, we apply extensive zero-padding to the truncated signal before performing the Fourier transform.

To quantify the wave packet dispersion, we characterize the localized spectrum $\tilde{\Psi}_n(\omega)$ using its energy-weighted central frequency $\omega_{c,n}$ and effective bandwidth variance $\sigma_{\omega,n}^2$ within a relevant physical band $[\omega_{\min}, \omega_{\max}]$
\begin{subequations}\label{eq:def-freq}
    \begin{equation}\label{eq:def-central-freq}
        \omega_{c,n} = \frac{\int_{\omega_{\min}}^{\omega_{\max}}\dif\omega\ \omega |\tilde{\Psi}_n(\omega)|^2}{\int_{\omega_{\min}}^{\omega_{\max}}\dif\omega\ |\tilde{\Psi}_n(\omega)|^2},
    \end{equation}
    \begin{equation}
        \sigma_{\omega,n}^2 = \frac{\int_{\omega_{\min}}^{\omega_{\max}} \dif\omega\  (\omega - \omega_{c,n})^2 |\tilde{\Psi}_n(\omega)|^2}{\int_{\omega_{\min}}^{\omega_{\max}}\dif\omega\ |\tilde{\Psi}_n(\omega)|^2}.
    \end{equation}
\end{subequations}
While $\omega_{c,n}$ generally differs from the absolute spectral peak $\omega_{{\rm peak},n}$, these two quantities coincide at leading-order for the broadband, single-peak signals analyzed in this work.

Applying this spectral analysis to the first echo at $\epsilon=10^{-5}$ yields the Fourier spectrum in Fig.\ \ref{fig:1-echo-FFT}. 
The distribution exhibits an extremely smooth, broadband single-peak structure, completely bypassing the discrete comb-like resonances predicted by the steady state pole condition $1-\mathcal{L}(\omega_n)=0$. 
The physical origin of this resonance absence is the system's extremely weak loop gain ($|R_{\rm bump}| \sim 10^{-5}$ at the fundamental frequency). 
Consequently, the single round-trip propagator satisfies $|\mathcal{L}(\omega_0)| \ll 1$. 
The trapped GW energy is dissipated into the horizon within a single period, plunging the cavity into a deeply over-damped regime. 
Without sufficient coherent energy to sustain multiple round-trip iterations, the formation of high-Q steady state resonances is physically prohibited.

Because the second echo is already severely distorted at $\epsilon=10^{-5}$, relying solely on this configuration is insufficient to extract the generic non-stationary evolution across multiple scatterings. 
We therefore scan the barrier strength across
\begin{equation}
    \epsilon \in \{10^{-6}, 10^{-5}, 10^{-4}, 10^{-3}\}.
\end{equation}
Here, $\epsilon=10^{-6}$ probes the deep over-damped limit, while $\epsilon=10^{-3}$ artificially enhances the late-time amplitude to provide a fully resolvable multi-echo sequence, allowing us to quantitatively track the dispersive trajectory of successive wave packets.

\begin{figure}[!htb]
    \centering
    \includegraphics[width=0.5\linewidth]{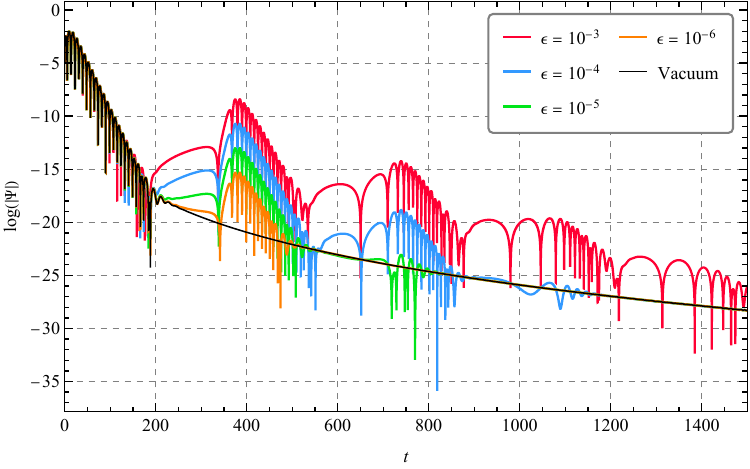}
    \caption{
    Logarithmic time domain waveforms $\log|\Psi(t)|$ for a fixed cavity length $L=180$ under varying barrier strengths $\epsilon \in \{10^{-6}, 10^{-5}, 10^{-4}, 10^{-3}\}$. 
    The black solid line represents the pure power law tail of the unperturbed vacuum Schwarzschild spacetime. 
    The divergence of the late-time waveforms explicitly demonstrates the dependence of the maximum viable echo number $n_{\max}$ on the barrier amplitude. 
    Notably, echoes are intrinsically truncated by the late-time power law tail, precluding the establishment of any steady state resonance in the low-finesse regime.}
    \label{fig:full-waveform}
\end{figure}

Two critical physical facts emerge from Fig.\ \ref{fig:full-waveform}.
First, at the extreme low-reflectivity limit ($\epsilon=10^{-6}$), single-trip energy leakage is so severe that the first echo is  partly submerged by the power law tail of prompt ringdown.
Second, for stronger perturbations ($\epsilon \geq 10^{-4}$), the system retains multiple resolvable echoes. These successive wave packets do not preserve the morphology of the prompt ringdown;
rather, they undergo progressive dispersive broadening and develop prominent asymmetric tails with increasing scattering order $n$.
By the third echo, the original wave packet structure is significantly distorted, with strong deformations in its low-frequency components.

\begin{figure}[!htb]
    \centering
    \begin{subfigure}[t]{0.5\textwidth}
         \centering
         \includegraphics[width=\textwidth]{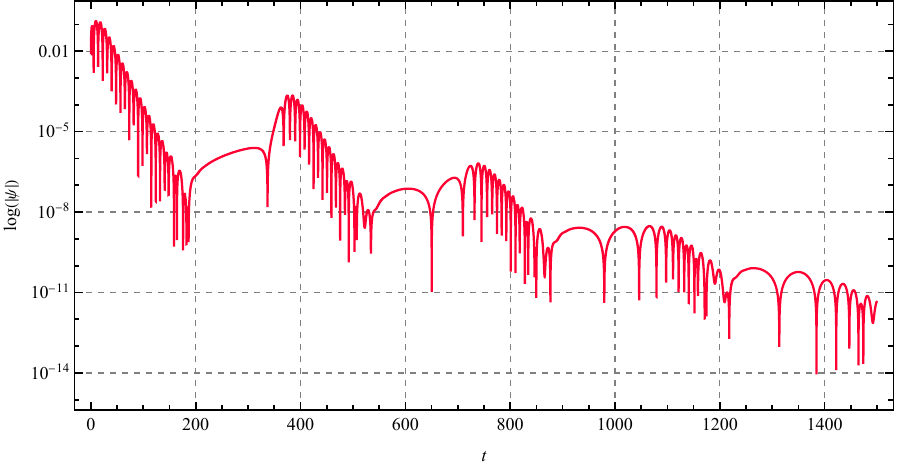}
         \caption{Logarithmic time domain waveform $\log|\Psi(t)|$, clearly exhibiting a sequence of echoes with progressive dispersive broadening.}
         \label{fig:3-echo-time}
     \end{subfigure}
     \begin{subfigure}[t]{0.405\textwidth}
         \centering
         \includegraphics[width=\textwidth]{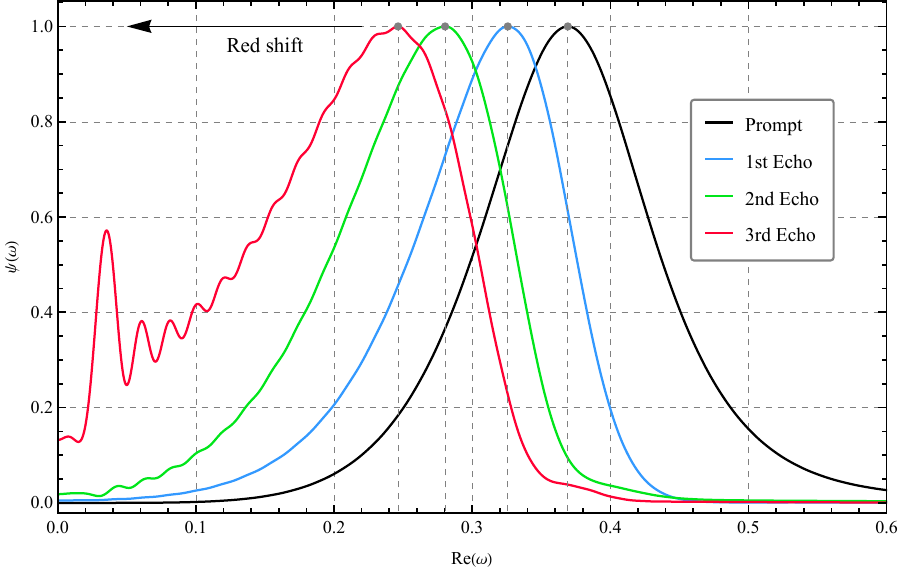}
         \caption{Fourier spectra of the prompt ringdown and the first three echoes, demonstrating the systematic redshift of the central frequency.}
         \label{fig:3-echo-frequency}
     \end{subfigure}
    \caption{Time-frequency evolution of the first three echoes for the amplitude-enhanced case ($\epsilon = 10^{-3}$).
    Here we adopt the time windows $t\in [0,360]$ for the prompt ringdown, $ [360,720]$ for the first echo, $ [720,1080]$ for the second echo, and $ [1080,1440]$ for the third echo. }
    \label{fig:3-echo}
\end{figure}

To diagnose the physical mechanism driving this morphological distortion, we analyze the amplitude-enhanced case ($\epsilon=10^{-3}$), which provides a resolvable multi-echo sequence. 
As shown in Fig.\ \ref{fig:3-echo}, while the first echo largely inherits the broadband spectrum of the prompt ringdown, successive round-trip scatterings induce a systematic and continuous redshift in its central frequency $\omega_{c,n}$.

This spectral drift is fundamentally governed by the iterative transfer function. 
The inner Regge-Wheeler barrier acts as a strongly frequency-dependent high-pass filter, preferentially transmitting high-frequency components into the horizon.
Consequently, the magnitude of the single-round-trip loop gain $|\mathcal{L}(\omega)|$ strictly satisfies
\begin{equation}
\frac{\dif}{\dif \omega} \log |\mathcal{L}(\omega)| < 0.
\end{equation}
Because the $n$-th echo spectrum scales as $|\tilde{\Psi}_n(\omega)| \simeq |\tilde{\Psi}_{\rm prompt}(\omega)||\mathcal{L}(\omega)|^n$, this negative gradient exponentially suppresses high-frequency components at higher scattering orders $n$. 
It is this asymmetric energy leakage that continuously drives the spectral centroid toward the low-frequency band.

Tab.\ \ref{tab:epsilon_scan} summarizes the core parameters of the wave packet evolution.
Data is extracted using fixed time windows for each event ($t \in [0, 360]$ for the prompt ringdown, $[360, 720]$ for the 1st echo, and $[720, 1080]$ for the 2nd echo).
The prompt ringdown consistently central frequency near $\omega_{c,p} \simeq \omega_{\rm peak} \simeq 0.374$, perfectly aligning with the fundamental $l=2$ Schwarzschild QNM $\omega_0 \simeq 0.374$.
It also shows that the central frequency of the wave packet coincides with the peak frequency at leading order.
Energetically, the amplitude ratio $A_1 / A_{\rm prompt}$ scales linearly with $\epsilon$. 
This ratio roughly approximates the main-band loop propagator $|\mathcal{L}(\omega_{\rm peak})| \ll 1$, quantitatively confirming the deeply over-damped nature of the system. 
Furthermore, in multi-echo sequences ($\epsilon \ge 10^{-4}$), the time interval between adjacent peaks deviates from the rigid geometric translation constant $2L = 360$ \cite{Abedi:2016hgu,Lo:2018sep,Uchikata:2019frs}, exhibiting a systematic time gliding (e.g., $\Delta t_1 \simeq 354.3$ at $\epsilon=10^{-3}$).

\begin{table}[!htb]
\centering
\caption{Phenomenological parameters extracted from the $\epsilon$ scan. Note that $A_{\rm prompt}$ and $A_n$ denote the absolute maximum amplitude of the prompt ringdown and the $n$-th resolvable echo, respectively. 
The energy-weighted central frequency $\omega_{c,n}$ and effective bandwidth $\sigma_{\omega,n}$ are computed within a physical frequency band $\omega \in [0.0, 0.6]$ using fixed time domain extraction windows (prompt: $t \in [0, 360]$, 1st echo: $t \in [360, 720]$, 2nd echo: $t \in [720, 1080]$). 
Notably, the unperturbed prompt ringdown exhibits a peak frequency $\omega_{\rm peak} \simeq 0.374$, which aligns with the fundamental $l=2$ Schwarzschild QNM.}
\label{tab:epsilon_scan}
\begin{tabular}{cccccccccc}
\hline\hline
$\epsilon$ & $n_{\max}$ & $t_1^{\rm peak}$ & $\Delta t_1$ ($n \ge 2$) & $\omega_{c,1}$ & $\sigma_{\omega,1}$ & $\omega_{c,2}$ & $\sigma_{\omega,2}$ & $A_1 / A_{\rm prompt}$ & $A_2/A_1$ \\
\hline
$10^{-6}$ & 1 & 674.7 & -- & 0.301 & 0.064 & -- & -- & $1.69 \times 10^{-6}$ & -- \\
$10^{-5}$ & 1 & 674.7 & -- & 0.301 & 0.064 & -- & -- & $1.69 \times 10^{-5}$ & -- \\
$10^{-4}$ & 2 & 674.7 & 354.5 & 0.301 & 0.064 & 0.255 & 0.072 & $1.69 \times 10^{-4}$ & $2.96 \times 10^{-4}$ \\
$10^{-3}$ & $\ge 3$ & 674.7 & 354.3 & 0.301 & 0.064 & 0.255 & 0.072 & $1.70 \times 10^{-3}$ & $2.95 \times 10^{-3}$ \\
\hline
\end{tabular}
\end{table}

Moreover, Tab.\ \ref{tab:epsilon_scan} quantifies the spectral drift induced by this non-stationary evolution. 
Between the first and second echoes, the central frequency $\omega_{c,n}$ redshifts from $0.301$ to $0.255$, while the effective bandwidth $\sigma_{\omega,n}$ broadens from $0.064$ to $0.072$. 
This low-frequency spectral asymmetry and bandwidth broadening directly correspond to the dispersion driven asymmetric tails observed in the time domain. 
These phenomenological contradictions unequivocally demonstrate the breakdown of the steady state picture, motivating the rigorous theoretical criteria developed in Sec.\ \ref{sec.III}.

\section{Breakdown of steady state resonance}\label{sec.III}

Phenomenology and QNM extraction (see App.\ \ref{app:A}) demonstrate that low-finesse environmental echoes lack resolvable steady state comb-like resonances. 
Although the global transfer function involves a resonant denominator $1/(1-\mathcal{L}(\omega))$ that mathematically generates a discrete family of cavity mode poles, these poles fail to manifest in the physical response. 
This section analytically establishes that the steady state resonance picture inevitably breaks down due to two independent constraints: 
\begin{enumerate}
    \item {\em Frequency domain spectral aliasing}:
    Extreme single-trip dissipation broadens the resonance linewidths beyond the free spectral range of adjacent poles.  
    This physically merges discrete peaks into a broadband dispersion-modulated signal.
    \item {\em Time domain power law tail truncation}: 
    The rapid exponential decay limits the viable echo sequence to $n_{\max} \sim \mathcal{O}(1)$ round trips before it is completely submerged by the late-time power law tail, physically precluding the coherent buildup required for standing waves.
\end{enumerate}

\subsection{Loop propagation and the transient nature of early echoes}

To track the dispersive evolution of the wave packet, we decompose the reflectivity and transmittance of the Regge-Wheeler and environmental barriers into their magnitudes and phases
\begin{equation}
R_X(\omega) = |R_X(\omega)| \mathrm{e}^{\mi\delta_X(\omega)}, \quad T_X(\omega) = |T_X(\omega)| \mathrm{e}^{\mi \varphi_X(\omega)},
\end{equation}
where $X \in \{\mathrm{RW}, \mathrm{bump}\}$. The frequency derivative $\delta_X^{\prime}(\omega) \equiv \dif\delta_X(\omega)/\dif\omega$ defines the Wigner time delay induced by reflection \cite{Wigner:1955zz}.
Unlike rigid boundaries where this delay is negligible, the strong phase nonlinearity of both the Regge-Wheeler and soft-environment barriers acts as the physical origin of the arrival-time gliding.

The loop propagator is thus rewritten as
\begin{equation}
\mathcal{L}(\omega) = |R_{\mathrm{RW}}(\omega) R_{\mathrm{bump}}(\omega)| \mathrm{e}^{\mi\Phi(\omega)},
\end{equation}
where the global phase accumulation $\Phi(\omega)$ encodes both the free propagation and the reflection phase shifts
\begin{equation}
    \Phi(\omega) = 2\omega L + \delta_{\mathrm{RW}}(\omega) + \delta_{\mathrm{bump}}(\omega).
\end{equation}
The corresponding group delay for a single round trip is
\begin{equation}
    T_{\mathrm{round}}(\omega) \equiv \frac{\dif\Phi(\omega)}{\dif\omega} = 2L + \delta^{\prime}_{\mathrm{RW}}(\omega) + \delta^{\prime}_{\mathrm{bump}}(\omega).
\end{equation}
Due to the strong dispersion of the soft barriers, $T_{\mathrm{round}}(\omega)$ dynamically deviates from the constant geometric time $2L$. 
This dispersive group delay strictly governs the systematic time gliding $\Delta t_n$ observed in the echo sequence.

The extremely small loop gain ($|\mathcal{L}(\omega)| \ll 1$) inevitably truncates the geometric series at a low order ($n \sim \mathcal{O}(1)$), completely masking the discrete mathematical poles and degenerating the frequency response into a continuous interference modulation signal.
To translate this phenomenological pole masking into a quantitative resolvability criterion, we bypass the construction of exact waveforms. 
Instead, we retain the frequency dependence of $\mathcal{L}(\omega)$ and weight-average it over the prompt ringdown spectrum $|\tilde{\Psi}_{\rm prompt}(\omega)|^2$. 
This yields the effective macroscopic parameters that characterize the net dissipation and characteristic lifetime of a single round trip.

\subsection{Resolvability bound from linewidth and free spectral range}

To establish a model-independent criterion for spectral aliasing, we invoke the Fourier-Gabor uncertainty principle.
For any physical signal $\Psi(t)$ with finite energy and its Fourier transform $\tilde{\Psi}(\omega)$, 
we define the effective temporal width of the signal as the second-order central moment of its time domain energy distribution,
\begin{subequations}
\begin{equation}
    \sigma_t^2 = \frac{\int \dif t\ (t-\langle t\rangle)^2 |\Psi(t)|^2}{\int \dif t \ |\Psi(t)|^2},
\end{equation}
where
\begin{equation}
    \langle t\rangle = \frac{\int \dif t\ t |\Psi(t)|^2}{\int \dif t\ |\Psi(t)|^2}
\end{equation}
\end{subequations}
is the temporal centroid of the signal energy distribution. 
Thus, $\sigma_t$ measures an effective duration of the signal rather than a dynamical decay constant.
With $\sigma_\omega$ defined analogously in the frequency domain, as shown in Eq.\ \eqref{eq:def-freq}. 
Two quantities satisfy the Fourier-Gabor inequality
\begin{equation}
    \sigma_t \sigma_\omega \geq \frac{1}{2}.
\end{equation}
This inequality implies that the shorter the duration of the time domain wave packet, the more severe its dispersion in the frequency domain must be.
In the over-damped environmental cavity, the rapid leakage of energy greatly compresses $\sigma_t$, which fundamentally prohibits the possibility of narrow peaks appearing in the frequency domain.

To translate this temporal constraint into a dynamical threshold, we account for the strong frequency dependence of the loop propagator. 
Since the $n$-th echo spectrum is modulated by $\mathcal{L}^n(\omega) = \exp[n \log \mathcal{L}(\omega)]$, we define a logarithmically weighted effective reflectivity $R_{A,\mathrm{eff}}$
\begin{equation}\label{eq:eff-reflectivity}
\log R_{A,\mathrm{eff}} \equiv \frac{\int_{\omega_{\min}}^{\omega_{\max}} \dif\omega\ W(\omega) \log |\mathcal{L}(\omega)| }{\int_{\omega_{\min}}^{\omega_{\max}} \dif\omega\ W(\omega)}.
\end{equation}
Here, the spectral weight $W(\omega) = |\tilde{\Psi}_{\rm prompt}(\omega)|^2$ is the prompt ringdown spectrum extracted at infinity, which appropriately averages the reflectivity over the dominant energy band ($0 < R_{A,\mathrm{eff}} < 1$).

We now map the discrete echo decay to a damped evolution process in continuous time.
For the peak evolution of the echo wave packets, their amplitudes approximately follow a geometric decay law
\begin{equation}
    A_n \simeq A_0 (R_{A,\mathrm{eff}})^n.
\end{equation}
Because the first echo carries the bulk of the trapped energy, we evaluate the frequency-dependent group delay at its central frequency $\omega_{c,1}$ to define a characteristic round-trip time $\bar{T}_{\mathrm{round}} \equiv T_{\mathrm{round}}(\omega_{c,1})$. 
This maps the discrete sequence to a continuous exponentially damped envelope
\begin{equation}
    A(t) \sim A_0 \me^{-\gamma t}, \quad \gamma = \frac{|\log R_{A,\mathrm{eff}}|}{\bar{T}_{\mathrm{round}}},
\end{equation}
with the macroscopic energy envelope decaying as $E(t) \propto A(t)^2 \sim \me^{-2 \gamma t}$.

In the complex frequency plane, the damping rate $\gamma$ corresponds to the imaginary part of the cavity poles, yielding a power spectrum with a Lorentzian profile $|\tilde{\Psi}(\omega)|^2 \propto [\gamma^2 + (\omega-\omega_0)^2]^{-1}$, where $\omega_0$ denotes the real part of the cavity modes. 
The full-width at half-maximum of this resonance is
\begin{equation}\label{eq:FWHM}
    \Gamma_{\mathrm{HWHM}} = \gamma, \quad \Gamma_{\mathrm{FWHM}} = 2\gamma = \frac{2|\log R_{A,\mathrm{eff}}|}{\bar{T}_{\mathrm{round}}}.
\end{equation}
The steady state coherent resonance condition $\Phi(\omega_m) = 2\pp m$ dictates the free spectral range between adjacent modes. 
Expanding the phase difference to first order yields
\begin{equation}\label{eq:FSR}
    \Delta\omega_{\mathrm{FSR}}(\omega_m) \equiv \omega_{m+1} - \omega_m \simeq \frac{2\pp}{\Phi^{\prime}(\omega_m)} = \frac{2\pp}{T_{\mathrm{round}}(\omega_m)}.
\end{equation}
Evaluating $\Delta\omega_{\mathrm{FSR}}$ at $\omega_{c,1}$, spectral aliasing physically occurs when the broadened resonant linewidth overlaps with adjacent peaks.  
We define the threshold for this breakdown as
\begin{equation}
\Gamma_{\mathrm{FWHM}} \gtrsim \eta \Delta\omega_{\mathrm{FSR}},
\end{equation}
where $\eta \sim \mathcal{O}(1)$ characterizes the spectral resolvability threshold. 
Substituting Eq.\ \eqref{eq:FWHM} and Eq.\ \eqref{eq:FSR} yields the theoretical bound
\begin{equation}
|\log R_{A,\mathrm{eff}}| \gtrsim \eta \pp.
\end{equation}
This inequality explicitly demonstrates that the observability of the comb-like spectrum depends entirely on the logarithmically weighted single-trip dissipation, and is fundamentally independent of the macroscopic cavity length $L$.

By substituting the specific value of the empirical parameter $\eta$, we present the corresponding upper threshold for the effective reflectivity.
For $\eta=1$, it corresponds to the aliasing threshold where the linewidth of adjacent peaks fully covers the peak spacing
\begin{equation}
    R_{A,\mathrm{eff}} \lesssim \mathrm{e}^{-\pp} \simeq 4.3 \times 10^{-2}.
\end{equation}
If we adopt the more conservative Rayleigh criterion with $\eta=2$ (i.e., the half-width at half-maximum satisfies $\Gamma_{\mathrm{HWHM}} \gtrsim \Delta\omega_{\mathrm{FSR}}$), we have
\begin{equation}
    R_{A,\mathrm{eff}} \lesssim \mathrm{e}^{-2\pp} \simeq 1.9 \times 10^{-3}.
\end{equation}
The above two expressions show that as long as the effective reflectivity is below the order of $\sim \mathcal{O}(10^{-2})$, the projection of the steady state poles onto the real frequency axis will inevitably undergo significant aliasing.
At this point, regardless of the value of the cavity length $L$ and how dense the poles of the transfer function are mathematically, the spectrum can only exhibit a smooth broadband envelope, without any resolvable steady state narrow-peak comb structure.

Although the numerical amplitude ratio $A_1 / A_{\rm prompt}$ extracted in Tab.\ \ref{tab:epsilon_scan} is not strictly identical to $R_{A,\mathrm{eff}}$, it serves as a robust order-of-magnitude estimate for the main-band single-loop gain. 
The extracted values ($A_1 / A_{\rm prompt} \le 1.70 \times 10^{-3}$ for $\epsilon \leq 10^{-3}$) explicitly confirm that all our low-finesse models fall deep within this spectral aliasing regime. 
This estimate is qualitatively consistent with Ref.\ \cite{Berti:2022xfj}, for very weak perturbations ($\epsilon \lesssim 10^{-3}$), cavity modes are not recoverable in the full waveform, whereas a clearly resonance-dominated late-time signal appears only for much stronger perturbations ($\epsilon \gtrsim 10^{-1}$).

\begin{figure}[!htb]
    \centering
    \includegraphics[width=0.5\linewidth]{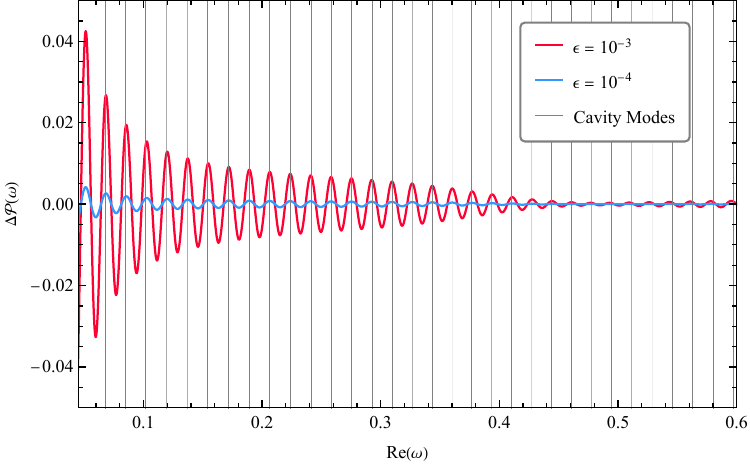}
    \caption{Prompt normalized spectrum for the enhanced models ($\epsilon=10^{-3}$ and $10^{-4}$). 
    The plotted quantity is $\Delta\mathcal{P}(\omega)\equiv |\tilde{\Psi}(\omega)|^2/|\tilde{\Psi}_{\rm prompt}(\omega)|^2-1$, evaluated over the main prompt-supported frequency band. 
    Instead of resolved narrow peaks, the spectra exhibit weak broadband oscillatory fringes, consistent with the cosine interference pattern predicted by the truncated scattering series in Eq.\ \eqref{eq:cos_modulation}. 
    Thin gray lines denote the theoretical resonant frequencies defined by $1-\mathcal{L}(\omega_n)=0$. 
    This illustrates that below the aliasing threshold, mathematical poles do not appear as resolvable steady-state cavity modes, but reduce to unresolvable interference modulations.}
    \label{fig:FFT-3-echos-comb}
\end{figure}

Within this aliased regime, the weak oscillatory structure exhibited by the prompt-normalized excess spectrum of the enhanced model (Fig.\ \ref{fig:FFT-3-echos-comb} for $\epsilon = 10^{-3}$) should not be misidentified as a set of resolved high-$Q$ steady-state cavity modes. 
To make this small interference pattern visible against the broad prompt envelope, we consider the dimensionless quantity
\begin{equation}\label{eq:deltaP_def}
\Delta \mathcal{P}(\omega)\equiv 
\frac{|\tilde{\Psi}(\omega)|^2}{|\tilde{\Psi}_{\rm prompt}(\omega)|^2}-1,
\end{equation}
evaluated over the main prompt-supported frequency band. 
In the low-reflectivity regime, truncating the scattering series at $N=1$ (retaining only the prompt ringdown and the first echo) gives
\begin{equation}
\tilde{\Psi}(\omega)\simeq \tilde{\Psi}_{\rm prompt}(\omega)\,[1+\mathcal{L}(\omega)] ,
\end{equation}
so that
\begin{equation}\label{eq:cos_modulation}
\Delta \mathcal{P}(\omega)\simeq 2\Re \mathcal{L}(\omega)
=2|\mathcal{L}(\omega)|\cos\Phi(\omega) .
\end{equation}
The oscillations therefore originate from broadband interference between the prompt ringdown and the first reflected component after truncation of the scattering series, rather than from a genuine steady-state resonance comb.
This confirms that these spectral oscillations are merely geometric cosine interference modulations between independent wave packets.  
They require neither steady state coherence nor resonant Lorentzian profiles. 
The manifestation of genuine resonant poles is strictly governed by the $\Gamma_{\mathrm{FWHM}}$ and $\Delta\omega_{\mathrm{FSR}}$ criterion;
their mere mathematical existence does not guarantee observable cavity modes in the time domain response (detailed in App.\ \ref{app:B}).

\subsection{Power law tail truncation and the loss of coherent buildup}

While the frequency domain criterion dictates spectral resolvability, establishing steady state resonance also demands sufficient time for coherent buildup. 
If the echo sequence decays below the late-time power law tail of prompt ringdown after only a few round trips, this coherent superposition is physically truncated.

Using the effective reflectivity defined in Eq.\ \eqref{eq:eff-reflectivity} above, the peak amplitude of the $n$-th echo can be rewritten as
\begin{equation}
    A_n \simeq A_0 \mathrm{e}^{-n|\log R_{A,\mathrm{eff}}|},
\end{equation}
where $A_0$ is the effective amplitude of the initial excitation wave packet of the system.

However, the BH spacetime fundamentally restricts the persistence of such ideal exponential decay at long time scales.
Since GWs inevitably undergo continuous long-range back-scattering by the curvature potential of the large-scale spacetime background during propagation, the perturbative signal will smoothly transition from exponential decay to a long-range power law tail at late evolutionary times.
The amplitude envelope of this background tail asymptotically satisfies
\begin{equation}
    A_{\mathrm{tail}}(t) \simeq C  t^{-p},
\end{equation}
where $C$ is a constant determined by the integral of the initial data.
The specific value of the power law exponent $p$ depends on the spin state of the BH and the multipole moment $l$ of the perturbation field;
for the axial gravitational perturbation in the Schwarzschild spacetime studied in this work, the decay exponent corresponding to the dominant quadrupole moment $l=2$ is theoretically predicted to be $p=2l+3=7$ \cite{Price:1971fb}.
To remain macroscopically observable, the $n$-th echo must exceed this background tail by a resolution threshold $\kappa \sim \mathcal{O}(1)$ at its arrival time $t_n \simeq n\bar{T}_{\mathrm{round}}$. 
This imposes a time domain balance equation for the maximum viable echo number $n_{\max}$
\begin{equation}
    A_0 \mathrm{e}^{-n_{\max}|\log R_{A,\mathrm{eff}}|} \simeq \kappa  C (n_{\max}\bar{T}_{\mathrm{round}})^{-p}.
\end{equation}
By introducing the decay rate $a \equiv |\log R_{A,\mathrm{eff}}|$ and the constant $Q \equiv A_0 (\bar{T}_{\mathrm{round}})^{p}/(\kappa C)$, this transcendental equation is recast into the standard form
\begin{equation}
    -\frac{a}{p} n_{\max} \mathrm{e}^{-\frac{a}{p}n_{\max}} = -\frac{a}{p} Q^{-1/p}.
\end{equation}
Solving via the Lambert $W$ function yields the exact analytical upper bound
\begin{equation}\label{eq:analytical-solution}
    n_{\max} = -\frac{p}{|\log R_{A,\mathrm{eff}}|} W_{-1} \left[ - \frac{|\log R_{A,\mathrm{eff}}|}{p} \left(\frac{\kappa C}{A_0 (\bar{T}_{\mathrm{round}})^{p}} \right)^{1/p} \right],
\end{equation}
The $W_{-1}$ branch is strictly selected because $n_{\max} > 0$ restricts the argument to the negative real axis. 
Since the exponential decay $|\log R_{A,\mathrm{eff}}|$ vastly outpaces the slowly varying power law tail, the logarithmic asymptotic behavior of $W_{-1}$ constrains $n_{\max}$ to an $\mathcal{O}(1)$ integer in the low-finesse regime.

We can verify this analytical bound against the numerical data in Tab.\ \ref{tab:epsilon_scan}. 
Fitting the unperturbed vacuum tail gives $C \simeq 3.26 \times 10^{10}$. 
Using the extracted parameters $A_0 \simeq 0.130$, $\bar{T}_{\mathrm{round}} \simeq 354.5$, and setting $\kappa = 2.0$, at the extreme low-finesse limit ($\epsilon=10^{-5}$), the decay parameter is $a \simeq 10.99$. 
Eq.\ \eqref{eq:analytical-solution} predicts $n_{\max} \simeq 1.58$. 
Since echoes are discrete events, this physically restricts the system to exactly $n_{\max}=1$ resolvable echo, matching our numerical extraction.
For the enhanced perturbation ($\epsilon=10^{-3}$), $a$ decreases to $6.38$, raising the analytical limit to $n_{\max} \simeq 3.64$. 
This accurately explains why exactly three viable wave packets can be resolved prior to truncation.

In summary, steady state resonance theory fundamentally fails in low-finesse macroscopic cavities due to two complementary constraints. 
In the frequency domain, extreme dissipation guarantees spectral aliasing, irrevocably smoothing out the discrete pole structure.
In the time domain, power law tail truncation physically prohibits the infinite round trips necessary for coherent standing-wave buildup.
Consequently, low-finesse early echoes are exclusively transient, finite-order scattering events. 
This fundamental breakdown necessitates abandoning steady state formalisms and constructing a non-stationary dispersive evolution theory purely based on the single-loop propagator $\mathcal{L}(\omega)$, which we develop in Sec.\ \ref{sec:IV}.

\section{Non-stationary dispersive dynamics of transient echoes}\label{sec:IV}

Having ruled out steady state resonances, the dynamics of low-finesse echoes are exclusively governed by the power law modulation of the prompt ringdown $\tilde{\Psi}_{\rm prompt}(\omega)$ by the loop propagator $\mathcal{L}^n(\omega)$.
This section establishes a unified analytical theory for this non-stationary dispersive evolution. 
Using solely the asymptotic prompt ringdown spectrum as input, we analytically derive the systematic redshift of the central frequency, the arrival time gliding, dispersive broadening, and the emergence of asymmetric tails. 
This framework completely abandons steady state assumptions, enabling direct theoretical predictions for the phenomenological parameters extracted in Tab.\ \ref{tab:epsilon_scan}.

We decouple the amplitude and phase of the $n$-th echo spectrum 
\begin{equation}
    \tilde{\Psi}_n(\omega) = \exp\left[P_n(\omega) + \mi\Theta_n(\omega)\right]
\end{equation}
where the logarithmic amplitude $P_n(\omega) \equiv \log|\tilde{\Psi}_n(\omega)|$ and phase $\Theta_n(\omega) \equiv \arg \tilde{\Psi}_n(\omega)$ scale exactly as
\begin{subequations}
\begin{equation}\label{eq:express-P}
    P_n(\omega) = \log|\tilde{\Psi}_{\mathrm{prompt}}(\omega)| + n \log|\mathcal{L}(\omega)|,
\end{equation}
\begin{equation}
    \Theta_n(\omega) = \arg\tilde{\Psi}_{\mathrm{prompt}}(\omega) + n \arg\mathcal{L}(\omega).
\end{equation}
\end{subequations}
In the over-damped limit, the environmental barrier imparts a strong frequency gradient to $|\mathcal{L}(\omega)|$. 
Crucially, the scattering order $n$ linearly amplifies the frequency derivatives of both $P_n(\omega)$ and $\Theta_n(\omega)$, driving the non-stationary drift of the wave packet's energy centroid and dispersive morphology after each round trip.

The $n$-th time domain echo is given by the inverse Fourier transform
\begin{equation}\label{eq:fourier-inverse}
\Psi_n(t) = \frac{1}{2\pp} \int_{-\infty}^{\infty}\dif \omega\ \exp \left[P_n(\omega) + \mi\Theta_n(\omega) - \mi \omega t\right].
\end{equation}
For localized transient wave packets, the integral is overwhelmingly dominated by the neighborhood of the spectral peak $\omega_{{\rm peak},n}$, where $P_n^{\prime}(\omega_{{\rm peak},n}) = 0$ and $P_n^{\prime\prime}(\omega_{{\rm peak},n}) < 0$. 
As noted in Sec.\ \ref{sec:II}, for such single-peaked narrowband signals, the energy-weighted central frequency $\omega_{c,n}$ coincides with this spectral peak at leading order ($\omega_{c,n} \simeq \omega_{{\rm peak},n}$). 
We therefore evaluate the maximum condition directly at $\omega_{c,n}$. 
Substituting Eq.\ \eqref{eq:express-P} yields the algebraic equation governing the spectral drift
\begin{equation}
P_{\mathrm{prompt}}^{\prime}(\omega_{c,n}) + n P_{\mathcal{L}}^{\prime}(\omega_{c,n}) = 0,
\end{equation}
where $P_{\mathrm{prompt}} \equiv \log|\tilde{\Psi}_{\mathrm{prompt}}|$ and $P_{\mathcal{L}} \equiv \log|\mathcal{L}(\omega)|$. Treating $\omega_{c,n}$ as an implicit continuous function of the scattering order $n$, implicit differentiation yields the analytical spectral drift rate
\begin{equation}\label{eq:spectra-drift}
\frac{\dif\omega_{c,n}}{\dif n} = -\frac{P_{\mathcal{L}}^{\prime}(\omega_{c,n})}{P_{\mathrm{prompt}}^{\prime\prime}(\omega_{c,n}) + n P_{\mathcal{L}}^{\prime\prime}(\omega_{c,n})}.
\end{equation}
The maximum condition guarantees a negative denominator. 
Consequently, the direction of the drift is dictated entirely by the numerator $P_{\mathcal{L}}^{\prime}(\omega_{c,n})$. Because the Regge-Wheeler barrier preferentially transmits high-frequency components into the horizon ($P_{\mathcal{L}}^{\prime}(\omega_{c,n}) < 0$), we have $\dif\omega_{c,n}/\dif n < 0$, analytically proving the systematic redshift observed in the early echoes.

We benchmark this analytical drift against the numerical $\epsilon=10^{-3}$ model from Tab.\ \ref{tab:epsilon_scan}. 
Evaluating the derivatives at the first echo's central frequency $\omega_{c,1} \simeq 0.301$ yields an instantaneous theoretical drift rate $\dif\omega_{c,n}/\dif n \simeq -0.044$. 
The corresponding discrete numerical redshift is $\Delta\omega_c = \omega_{c,2} - \omega_{c,1} \simeq -0.046$. 
Although the continuous derivative $\dif\omega_{c,n}/\dif n$ naturally deviates from the macroscopic finite-difference step $\Delta n = 1$ (further compounded by time-windowing effects), their consistent strict negativity and close numerical agreement firmly validates Eq.\ \eqref{eq:spectra-drift} as the dynamical origin of the low-finesse spectral redshift.

To analytically reconstruct the temporal distortion of the echo wave packet via Eq.\ \eqref{eq:fourier-inverse}, we perform a local Taylor expansion of $P_n(\omega)$ and $\Theta_n(\omega)$ around $\omega_{c,n}$.
Defining $\Delta\omega \equiv \omega - \omega_{c,n}$, we expand the amplitude to second order and the phase to third order.
The logarithmic amplitude yields a Gaussian profile
\begin{equation}\label{eq:series-P}
P_n(\omega) \simeq P_n(\omega_{c,n}) - \frac{\Delta\omega^2}{2\sigma_{\omega,n}^2},
\end{equation}
where the effective frequency domain bandwidth $\sigma_{\omega,n}$ satisfies
\begin{equation}
\sigma_{\omega,n}^{-2} \equiv -P_n^{\prime\prime}(\omega_{c,n}) = - P_{\mathrm{prompt}}^{\prime\prime}(\omega_{c,n}) - n P_{\mathcal{L}}^{\prime\prime}(\omega_{c,n}) > 0.
\end{equation}
This explicitly demonstrates that the progressive accumulation of the loop propagator's second derivative $P_{\mathcal{L}}^{\prime\prime}$ drives the steady bandwidth broadening observed in Tab.\ \ref{tab:epsilon_scan}.

A subtle point is that, the amplitude expansion $P_n(\omega)$ must be strictly truncated at second order. 
Retaining a cubic term $P_n^{\prime\prime\prime}(\omega_{c,n})\Delta\omega^3 / 6$ inside the real exponential would inevitably cause the inverse Fourier integral to diverge as $\Delta\omega \to \pm\infty$, violating the finite-energy condition of the time domain signal.  
Therefore, the second-order Gaussian suppression kernel is mathematically mandated; 
any higher-order amplitude skewness must be factored out of the exponential as polynomial corrections. 
This rigorous Edgeworth-type perturbative treatment is detailed in App.\ \ref{app:C}.

While amplitude expansion dictates integral convergence, the wave packet's dispersion is strictly governed by the phase $\Theta_n(\omega)$. 
To analytically capture the dispersion driven asymmetric tails, we expand the phase to third order around $\omega_{c,n}$
\begin{equation}\label{eq:series-Q}
\Theta_n(\omega) \simeq \Theta_n(\omega_{c,n}) + t_n\Delta\omega + \frac{D_n}{2}\Delta\omega^2 + \frac{K_n}{6}\Delta\omega^3,
\end{equation}
defining the temporal dispersion parameters
\begin{equation}
t_n \equiv \Theta_n^{\prime}(\omega_{c,n}), \quad D_n \equiv \Theta_n^{\prime\prime}(\omega_{c,n}), \quad K_n \equiv \Theta_n^{\prime\prime\prime}(\omega_{c,n}).
\end{equation}
Physically, the first derivative $t_n$ determines the group arrival time;
the second-order dispersion $D_n$ induces a linear frequency chirp and symmetric envelope broadening;
and the third-order dispersion $K_n$ explicitly breaks the symmetry, generating long-tailed asymmetric distributions.

Substituting Eqs.\ \eqref{eq:series-P} and \eqref{eq:series-Q} into Eq.\ \eqref{eq:fourier-inverse}, the integrand elegantly factors into a quadratic and a cubic exponential kernel. 
Defining the constant amplitude factor $\mathcal{A}_n \equiv \exp[P_n(\omega_{c,n})]$ and the complex broadening parameter $Q_n \equiv \sigma_{\omega,n}^{-2} - \mi D_n$, the inverse Fourier transform of the purely quadratic kernel yields a standard chirped Gaussian,
\begin{align}
g_n(\tau) &= \frac{1}{2\pp} \int_{-\infty}^{\infty} \dif \Delta\omega \ \exp \left[ -\frac{1}{2}Q_n \Delta\omega^2 + \mi \tau \Delta\omega \right] \nonumber \\
&= \frac{1}{\sqrt{2\pp Q_n}} \exp \left[ -\frac{\tau^2}{2Q_n} \right],
\end{align}
where $\tau = t_n-t$. 
The real part of $Q_n^{-1}$ governs the symmetric broadening, while its imaginary part drives the linear phase chirp.

The cubic kernel entirely embodies the third-order phase dispersion. 
Introducing the characteristic time scale $\ell_n \equiv (|K_n|/2)^{1/3}$, its inverse transform maps directly to the Airy function
\begin{align}
a_n(\tau) & = \frac{1}{2\pp} \int_{-\infty}^{\infty} \dif \Delta\omega \ \exp \left[ \mi \frac{K_n}{6}\Delta\omega^3 - \mi \tau\Delta\omega \right] \nonumber \\
& = \frac{1}{\ell_n} \mathrm{Ai} \left(-\frac{\tau}{\ell_n}\right).
\end{align}
This kernel constitutes the fundamental mathematical origin of symmetry breaking. 
The Airy function $\mathrm{Ai}(x)$ decays exponentially on the positive half-axis but exhibits dense, slowly decaying oscillations on the negative half-axis, natively inducing the characteristic asymmetric tails.

The terminal morphology of the wave packet hinges on the accumulated third-order phase within the effective bandwidth. 
In the weak dispersion regime ($|K_n| \sigma_{\omega,n}^3 \ll 1$), the quadratic kernel $g_n(\tau)$ dominates, maintaining a symmetric chirped Gaussian profile. 
However, successive round-trip scatterings progressively amplify $K_n$. 
Once the system enters the strong dispersion regime ($|K_n| \sigma_{\omega,n}^3 \gtrsim 1$), rigid translations and pure Gaussian approximations fundamentally fail.

The complete non-stationary evolution must then be rigorously described by the time domain convolution of $g_n(\tau)$ and $a_n(\tau)$
\begin{equation}\label{eq:echo-reconstruct}
\Psi_n(t) \simeq \mathcal{A}_n \exp \left[\mi \left(\Theta_n(\omega_{c,n})-\omega_{c,n} t\right)\right] \int_{-\infty}^{+\infty} \dif u\  \mathrm{Ai} \left(\frac{u-(t-t_n)}{\ell_n}\right) \exp \left[-\frac{u^2}{2Q_n}\right].
\end{equation}
This analytical convolution elegantly merges the symmetric broadening and chirp of the complex Gaussian with the severe asymptotic asymmetry of the Airy function. 
It natively preserves the main energy peak while dragging out prominent asymmetric tails, precisely reproducing the temporal distortion of the late-time echoes observed in Fig.\ \ref{fig:3-echo}.

\section{Template Mismatch and the Dynamic Dispersive Template}\label{sec:V}

Sec.\ \ref{sec:IV} established that low-finesse echoes are not simple time domain translations, but are governed by the non-stationary co-evolution of $\{\omega_{c,n}, \sigma_{\omega,n}, t_n, D_n, K_n\}$.
Consequently, any viable waveform template must dynamically update these parameters with the scattering order $n$. 
Traditional rigid translation templates inherently suffer from a fundamental structural mismatch with the underlying dispersive dynamics. 
This section quantifies this structural failure and validates the reconstruction fidelity of our dynamic analytical template Eq.\ \eqref{eq:echo-reconstruct} against the exact numerical transfer function benchmark.

To quantify template fidelity, we employ the standard frequency domain inner product of two given time domain waveforms $\Psi_1$ and $\Psi_2$ over the relevant physical band $[\omega_{\min}, \omega_{\max}]$
\begin{equation}
\langle \Psi_1|\Psi_2\rangle = 4 \Re \int_{\omega_{\min}}^{\omega_{\max}} \dif \omega \ \tilde{\Psi}_1(\omega)\tilde{\Psi}_2^*(\omega).
\end{equation}
The normalized matching degree $\mathcal{M}$ between the exact numerical signal $\Psi_{\mathrm{true}}$ and the analytical template $\Psi_{\mathrm{temp}}$ is obtained by maximizing over the arrival time $t_c$ and global phase $\phi_c$
\begin{equation}\label{eq:match_def}
\mathcal{M} = \max_{t_c,\phi_c} \frac{\langle \Psi_{\mathrm{true}}|\Psi_{\mathrm{temp}}(t_c,\phi_c)\rangle}{\sqrt{\langle \Psi_{\mathrm{true}}|\Psi_{\mathrm{true}}\rangle\langle \Psi_{\mathrm{temp}}|\Psi_{\mathrm{temp}}\rangle}}.
\end{equation}

The traditional rigid translation template assumes fixed amplitude decay $q$ and constant geometric delay $T_0 = 2L$
\begin{equation}
    \Psi_{\mathrm{rigid}}(t) = \sum_{n=0}^{n_{\max}} q^n \Psi_{\rm prompt}(t-n T_0),
\end{equation}
where $q \in (0,1)$ denotes the fixed decay factor of amplitude. 
In a low-finesse cavity, this mathematically rigid assumption faces two inescapable structural mismatch mechanisms:
\begin{enumerate}
    \item {\em Phase decoherence from time gliding}: 
    The true arrival time $t_n$ deviates from $n T_0$ due to boundary dispersion.
    This time gliding $\Delta t_n \equiv t_n - n T_0$ introduces a phase deviation $\exp(-\mi \omega \Delta t_n)$ in the frequency domain. For a wave packet with finite bandwidth $\sigma_{\omega,n}$, this directly induces integral decoherence, suppressing the matching by an exponential factor
    \begin{equation}
    \sim \exp\left[ -\frac{1}{4}\sigma_{\omega,n}^2(\Delta t_n)^2 \right].
    \end{equation}
    Once the effective phase deviation $\sigma_{\omega,n}\Delta t_n$ accumulates to $\mathcal{O}(1)$, the matching degree collapses exponentially.
    \item {\em Spectral mismatch from frequency drift}:
    The rigid template forcibly assumes a constant carrier frequency $\omega_{c,0}$. 
    However, Eq.\ \eqref{eq:spectra-drift} dictates a continuous redshift. 
    The spectral deviation $\Delta\omega_n \equiv \omega_{c,n} - \omega_{c,0}$ exponentially suppresses the frequency domain overlap integral by
    \begin{equation}
    \sim \exp\left[ -\frac{(\Delta\omega_n)^2}{4\sigma_{\omega,n}^2} \right].
    \end{equation}
    When the spectral drift $\Delta\omega_n$ approaches the local bandwidth $\sigma_{\omega,n}$, the rigid template and the true dispersive signal become fundamentally orthogonal in the Hilbert space.
\end{enumerate}

Accordingly, we construct a dynamic analytical template by superimposing the non-stationary dispersive wave packets derived in Eq.\ \eqref{eq:echo-reconstruct}
\begin{equation}\label{eq:dynamic-template}
\Psi_{\mathrm{dyn}}(t) = \sum_{n=1}^{n_{\max}} \Psi_n(t),
\end{equation}
where $n_{\max}$ is the maximum observable echo number dictated by the power law tail truncation. 
Unlike rigid translations, each $\Psi_n(t)$ here represents a dynamically updated, non-stationary scattering event.

Because the prompt ringdown, first echo, and late-time tail cannot be separated sharply in the time domain, finite time-windowing and discrete sampling intrinsically cap the maximum achievable matching degree below $\mathcal{M}=1$.
Using the complete, unapproximated loop propagator, we construct the full transfer-function first-echo signal $\tilde{\Psi}_{\mathrm{exact}}(\omega) = \tilde{\Psi}_{\mathrm{prompt}}(\omega) \mathcal{L}(\omega)$. 
The inverse Fourier transform of this signal establishes the relevant benchmark for the analytical Taylor reconstruction within the same extraction framework.

\begin{figure}[!htb]
    \centering
    \includegraphics[width=0.6\linewidth]{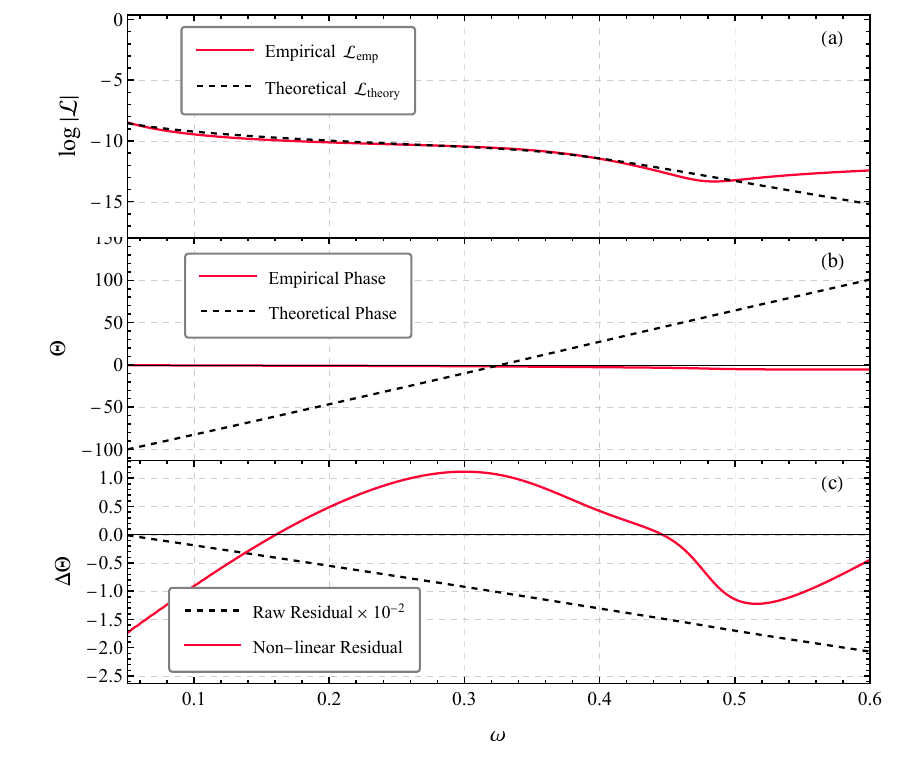}
    \caption{Comparison between the theoretical one-loop propagator $\mathcal{L}(\omega)$ and its empirical extraction for the $\epsilon=10^{-5}$ model over the analysis band $\omega \in [0.0, 0.6]$. 
    (a) Logarithmic magnitude: empirical (red solid) versus theoretical (black dashed). The agreement is best in the central prompt-supported region around the dominant saddle frequency, while visible deviations develop toward the upper band edge. 
    (b) Complex phase: empirical (black solid) versus theoretically aligned phase (red dashed), anchored at $\omega_{c,1}$. 
    (c) Phase residual: raw residual (black dashed) and the nonlinear residual (red solid) after subtracting the optimal linear time-delay trend. The remaining residual is smooth and structured rather than vanishing identically, indicating that the theoretical propagator captures the leading dispersive phase evolution but does not reproduce the empirical phase pointwise across the full band.}
    \label{fig:tf_diagnostics}
\end{figure}

To rigorously validate the theoretical one-loop propagator $\mathcal{L}_{\mathrm{theory}}(\omega)$—which is synthesized by multiplying the individual complex reflectivities of the Regge-Wheeler and bump potentials obtained via independent frequency scans—we benchmark it against an empirical propagator $\mathcal{L}_{\mathrm{emp}}(\omega) \equiv \tilde{\Psi}_{1}(\omega) / \tilde{\Psi}_{\mathrm{prompt}}(\omega)$ extracted directly from the realistic low-finesse numerical time domain spectra ($\epsilon=10^{-5}$). 
Fig.\ \ref{fig:tf_diagnostics} illustrates this critical validation.

As shown in Fig.\ \ref{fig:tf_diagnostics}(a), the theoretical amplitude reproduces the main trend of the empirically extracted one-loop response in the dominant prompt-supported band, especially around the saddle region that controls the subsequent template construction, although noticeable deviations appear toward the upper band edge. 
In panels (b) and (c), after anchoring the global phase at $\omega_{c,1}$ and subtracting the optimal linear phase trend $\omega \Delta t$ associated with the macroscopic propagation delay, the remaining phase mismatch is reduced to a smooth nonlinear residual. 
This shows that the exact theoretical propagator $\mathcal{L}_{\mathrm{theory}}(\omega)$ captures the leading local scattering dynamics relevant for the dominant echo band, but should be interpreted as a benchmark for subsequent template matching rather than an exact reconstruction of the empirically extracted transfer function over the entire plotted interval.

\begin{figure}[!htb]
    \centering
    \includegraphics[width=0.5\linewidth]{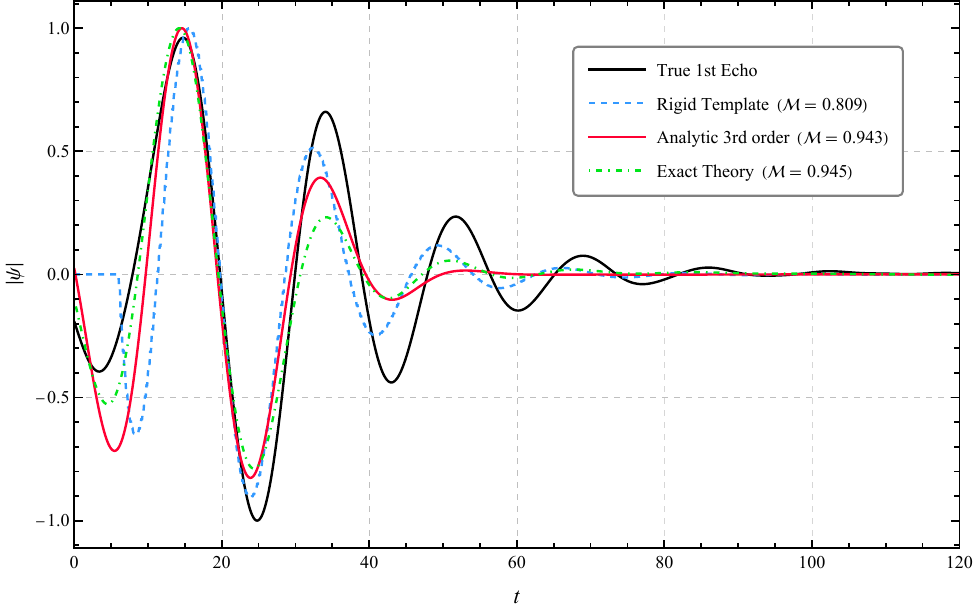}
    \caption{Time domain reconstruction of the first echo ($\epsilon= 10^{-5}$).
    The true 1st echo signal (black solid) is compared against the exact transfer function benchmark (green dash-dotted), the dynamic dispersive template (red solid), and the traditional rigid translation template (blue dashed).
    Matching degrees are maximized over $(t_c, \phi_c)$ within $\omega \in [0.0, 0.6]$. 
    The severe temporal and morphological mismatch of the rigid template visually confirms the necessity of incorporating non-stationary dispersive parameters.}
    \label{fig:reconstruct}
\end{figure}

A comparison between the original first echo signal and the signals reconstructed by the rigid template, the analytical template given by Eq.\ \eqref{eq:echo-reconstruct}, and the full transfer function is shown in Fig.\ \ref{fig:reconstruct}.
After maximizing the parameters $(t_c, \phi_c)$, the full transfer-function reconstruction yields a matching degree of $\mathcal{M}_{\mathrm{exact}} \simeq 0.945$ with respect to the FDTD-extracted first echo.
Here $\mathcal{M}_{\mathrm{exact}}$ provides the benchmark for the present linear scattering extraction. 
Applying the Taylor expanded dispersive reconstruction to the same first echo packet gives $\mathcal{M}_{\mathrm{ana}} \simeq 0.943$, only $\Delta \mathcal{M}\simeq 0.002$ below this full transfer function benchmark.
Thus the Taylor reconstruction loses very little overlap relative to the exact transfer function reconstruction within the same extraction framework. 
This small loss shows that the parameter set $\{\omega_{c,n}, \sigma_{\omega,n}, t_n, D_n, K_n\}$ captures the dominant morphology of the first echo packet generated by the full transfer function in the analyzed frequency band.

In the context of precision parameter estimation, an absolute mismatch at the level of $1-\mathcal{M}_{\mathrm{ana}}\simeq 5.7\%$ would still be considered sizable. 
As anticipated above, this absolute mismatch is affected by the finite time domain extraction and Fourier-windowing procedure, because the first echo cannot be cleanly separated from the prompt ringdown tail and the subsequent backscattering response. 
It should therefore not be identified with the physical error introduced by the Taylor expansion alone, nor should it be directly compared with the precision mismatches quoted for numerical relativity calibrated binary BH waveform models. 
Determining the recovery accuracy of the morphology parameters $\{\omega_{c,n},\sigma_{\omega,n},t_n,D_n,K_n\}$, including possible systematic biases, would require dedicated injection-recovery studies. 
Within the present linear scattering framework, such studies should use the full FDTD waveforms as target signals against which any reduced analytical template is calibrated. 
However, because echo modeling still lacks a fully nonlinear numerical relativity waveform, this validation would remain internal to the chosen linear model and should not be extrapolated to detector level search efficiencies or model independent parameter estimation accuracies.

\section{Conclusion and discussion}\label{sec:VI}

In this work, we established a non-stationary dispersive evolution theory for GW echoes in low-finesse environments. 
By identifying two independent physical thresholds—frequency domain spectral aliasing and time domain power law tail truncation, we quantitatively delineated the breakdown of the steady state resonance picture. 
Within this breakdown regime, early echoes are strictly transient scattering events governed by a unified 5-parameter set $\{\omega_{c,n}, t_n, \sigma_{\omega,n}, D_n, K_n\}$. 
This framework natively captures the central frequency redshift, arrival time gliding, dispersive broadening, and asymmetric tails; 
for the first echo, its overlap with the FDTD-extracted waveform differs from the full transfer-function reconstruction by only $\Delta\mathcal{M} \simeq 0.002$ within the same finite-window extraction.
Consequently, Eq.\ \eqref{eq:echo-reconstruct} is not a mere phenomenological fit, but the fundamental analytical form of localized dispersive echoes. 
It conceptually separates the mere ``mathematical existence of poles" from the ``physical observability of standing waves", placing low-finesse echo dynamics on a strictly computable and interpretable foundation.

The non-rigid, dispersive nature of GW echoes has been progressively recognized across distinct theoretical frontiers. 
Early time domain studies of exotic compact objects demonstrated that late-time signals intrinsically manifest as modulated, distorted pulse trains rather than rigid replicas of the prompt ringdown \cite{Cardoso:2016rao,Nakano:2017fvh}. 
Formally, Green's function approaches have elegantly interpreted these early echoes as reprocessed responses of the perturbation propagator \cite{Mark:2017dnq,Correia:2018apm}. 
In spinning exotic compact object spacetimes, analytic templates explicitly incorporate frequency-dependent barrier filtering to capture the progressive spectral evolution of successive pulses \cite{Maggio:2019zyv}. 
Furthermore, broader phenomenological studies have rightfully relaxed the assumption of strictly equidistant echo spacing \cite{Wang:2018mlp}. 
Our work analytically unifies these complementary insights. 
By rigorously defining the exact thresholds where the steady state resonance picture collapses, we provide the fundamental non-stationary dispersive dynamics that universally govern this transient evolution in low-finesse environments.

This fundamental physical paradigm shift carries direct strategic implications for GW data analysis.
Traditional searches targeting uniform, phase-coherent comb spectra \cite{Conklin:2017lwb,Ren:2021xbe} are intrinsically structurally mismatched with realistic low-finesse signals, which undergo continuous phase-space drift. 
Our five-parameter model bridges the gap between full-numerical models and weak-model searches by providing a compact morphology-level description of the FDTD echo segments. 
For matched-filter searches, the conservative strategy is to use echo segments generated by the full FDTD evolution or by a strict numerical transfer-function calculation as templates, while using $\{\omega_{c,n}, t_n, \sigma_{\omega,n}, D_n, K_n\}$ to interpret, interpolate, and organize their non-stationary morphology. 
Conversely, in the absence of precise priors, proving this inevitable non-stationary drift fundamentally justifies the adoption of morphology-relaxed time-frequency search strategies. 
Algorithms such as the Coherent WaveBurst or wavelet-based methods \cite{Tsang:2018uie,Miani:2023mgl} are natively better equipped to capture transient pulse trains that continuously distort and migrate across the time-frequency plane.

Finally, the local dispersion theory developed herein delineates clear pathways for future extensions. 
While our single-center framework accurately captures localized broadband spectra, highly multi-peaked ringdown signals would naturally necessitate a generalization to multi-center interference. 
This work utilizes the Schwarzschild spacetime as a pristine theoretical laboratory, strictly isolating the intrinsic dispersive dynamics driven by boundary potentials. 
In realistic astrophysical environments, rotating BHs introduce profound spin-dependent modifications that fundamentally alter the boundary reflection and transmission properties \cite{Nakano:2017fvh,Maggio:2019zyv}. 
While our current analysis assumes a strictly stationary and spherically symmetric exterior, remnants of neutron star mergers can involve nonlinear, time-dependent fluid environments where transient, matter-driven oscillations mimic echo-like structures without a fixed external barrier \cite{Steppohn:2025kbh}.
For such a genuinely dynamical background, time-translation invariance is broken.
Consequently, the standard algebraic relation $\Psi_{\rm out}(\omega)=\mathcal{L}(\omega)\Psi_{\rm in}(\omega)$ is no longer valid.
The response must instead be formulated in the time domain using a causal two-time scattering kernel, $\Psi_{\rm out}(t)=\int \dif t^{\prime}\ \mathcal{K}(t,t^{\prime})\Psi_{\rm in}(t^{\prime})$, where $t^{\prime}$ is the injection time and $t$ is the observation time.
In frequency domain, this induces an intricate coupling among distinct frequency components.
Therefore, the present framework is restricted to stationary configurations and cannot describe the time-dependent fluid environments of merger remnants.
Before tackling fully dynamical fluids, self-consistent thin-shell models \cite{Laeuger:2025zgb} provide a rigorous analytical foundation, demonstrating how dynamical junction conditions preserve the prompt vacuum ringdown while generating weak late-time echoes.
Extending this non-stationary dispersive framework to Kerr spacetimes, thereby quantifying the intricate interplay between BH spin and macroscopic environmental dispersion, remains the paramount next step.

\appendix
\section{Quasinormal mode extraction from the first echo}\label{app:A}

This appendix provides independent methodological verification for the breakdown of the steady state resonance picture via QNM extraction.

If the first echo had established a steady state resonance, its frequency would be strictly quantized by the global pole condition $1 - \mathcal{L}(\omega_n) = 0$.
Parameterizing the boundary reflections as $R_X(\omega) = |R_X(\omega)| \me^{\mi \delta_X(\omega)}$, this necessitates the standing-wave phase condition
\begin{equation}
2\omega_n L + \delta_{\rm tot}(\omega_n) = 2\pp n, \quad (n \in \mathbb{Z}),
\end{equation}
where $\delta_{\rm tot} \equiv \delta_{\mathrm{RW}} + \delta_{\mathrm{bump}}$ is the total round-trip reflection phase shift. 
Perturbing around the purely geometric eigenfrequency $\omega_n^{(0)} = n \pp /L$ (letting $\omega_n = \omega_n^{(0)} + \Delta\omega_n$), the steady state picture analytically predicts a boundary-induced frequency shift
\begin{equation}
\Delta\omega_n \simeq -\frac{\delta_{\rm tot}(\omega_n)}{2L}.
\end{equation}
Evaluating this at the dominant Schwarzschild fundamental frequency $\omega_0 \simeq 0.374$ yields $\delta_{\mathrm{RW}} \simeq -1.07\ \mathrm{rad}$ and $\delta_{\mathrm{bump}} \simeq -1.57\ \mathrm{rad}$. The strictly negative total phase shift $\delta_{\mathrm{tot}} \simeq -2.64\ \mathrm{rad}$ dictates a definitive theoretical prediction—if the first echo is a steady state cavity mode, its central frequency must exhibit a macroscopic blue shift scaling as $1/L$.

To quantitatively determine the exact complex frequencies of these hypothetical modes, we directly compute the mathematical poles for the full perturbed potential $V_{\rm RW}+V_{\rm bump}$. 
We impose the standard QNM boundary conditions
\begin{equation}
    \Psi\sim \me^{-\mi\omega r_*}\ (r_*\to-\infty), \quad \Psi\sim \me^{+\mi\omega r_*}\ (r_*\to+\infty),
\end{equation}
and determine the complex poles via a two-sided shooting calculation. 
The ingoing solution is initialized by a Frobenius expansion near the horizon, and the outgoing solution by an asymptotic expansion at large radius. 
The eigenvalue condition is imposed by requiring the two independently integrated solutions to be linearly dependent at the matching point $r_m=6M$, or equivalently by the vanishing of their Wronskian there.
After fixing their relative normalization by $b=\Psi_{\rm H}(r_m)/\Psi_\infty(r_m)$, we solve the normalized derivative matching residual
\begin{equation}
    \mathcal{J}(\omega)\equiv \frac{\Psi_{\rm H}^{\prime}(r_m)-b\Psi_\infty^{\prime}(r_m)}{\Psi_{\rm H}^{\prime}(r_m)}=0.
\end{equation}
Tab.\ \ref{tab:qnm_shooting} presents the complex poles for $\epsilon=10^{-3}$ case. 
The results show that the mathematical poles indeed form the expected cavity ladder, with a real-frequency spacing approaching the geometric estimate $\pp/L \simeq 1.75 \times 10^{-2}$, which is consistent with the steady-state phase condition.
However, the exact calculation also reveals a critical physical limitation: even at this critical amplitude, the damping width of these modes ($2|\Im\omega_n|$) reaches approximately $1.6 \sim 1.8$ times the adjacent spacing.
This severe spectral aliasing fundamentally guarantees that these discrete poles cannot manifest as a resolved high-$Q$ comb in the low-finesse early time domain response.
\begin{table}[!htb]
\centering
\caption{Representative cavity poles for the marginal critical double-barrier model with $l=2$, $L=180$, and $\epsilon=10^{-3}$.
The last column compares the pole full width $2|\Im\omega_n|$ with the adjacent real-frequency spacing $\Delta\omega$.}
\label{tab:qnm_shooting}
\begin{tabular}{lccc}
\hline\hline
$n$ & $\Re(\omega_n)$ & $\Im(\omega_n)$ & $2|\Im\omega_n|/\Delta\omega$\\
\hline
5  & 0.083262 & -0.012595 & 1.44\\
6  & 0.101115 & -0.013069 & 1.49\\
7  & 0.118841 & -0.013473 & 1.54\\
8  & 0.136474 & -0.013826 & 1.58\\
9  & 0.154034 & -0.014138 & 1.62\\
10 & 0.171531 & -0.014418 & 1.65\\
11 & 0.188971 & -0.014672 & 1.68\\
12 & 0.206358 & -0.014904 & 1.70\\
13 & 0.223692 & -0.015117 & 1.73\\
14 & 0.240972 & -0.015314 & 1.75\\
\hline\hline
\end{tabular}
\end{table}

\begin{figure}[!htb]
    \centering
    \begin{subfigure}[t]{0.45\textwidth}
         \centering
         \includegraphics[width=\textwidth]{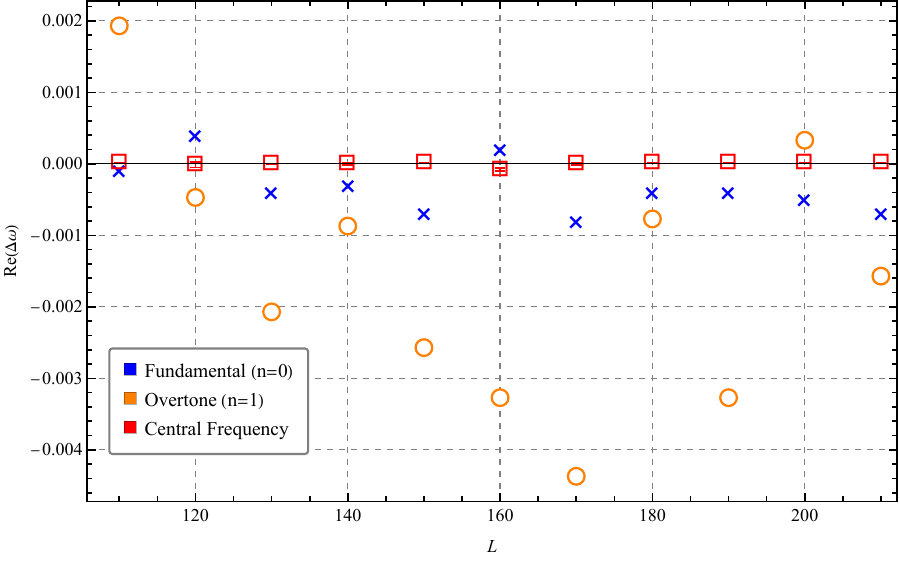}
         \caption{Real part frequency shift $\Re(\Delta \omega)$. The central frequency strictly locks to the unperturbed baseline ($\Delta \omega \simeq 0$), definitively rejecting the $1/L$ blue-shift prediction of the steady state model.}
         \label{fig:deviation-re}
     \end{subfigure}
     \begin{subfigure}[t]{0.45\textwidth}
         \centering
         \includegraphics[width=\textwidth]{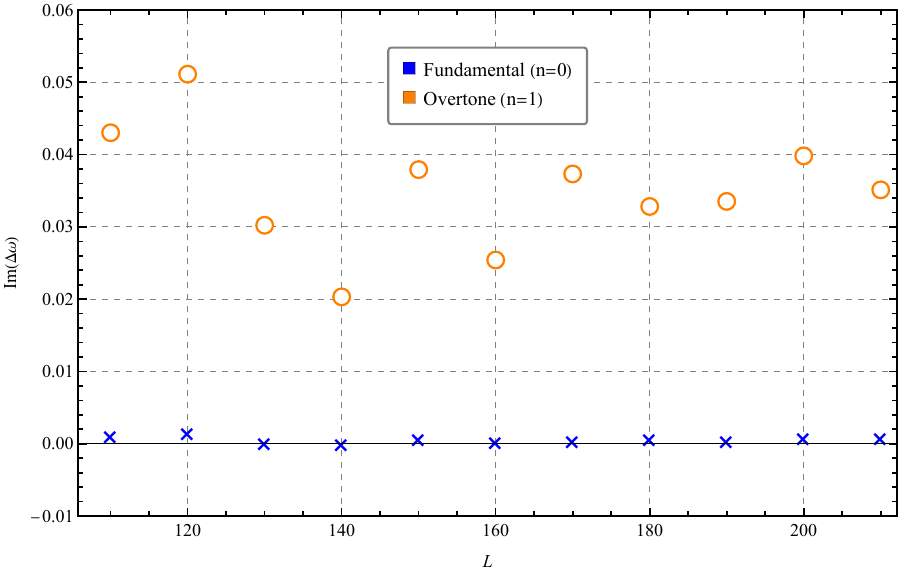}
         \caption{Imaginary part shift $\Im(\Delta \omega)$. The fundamental mode (blue crosses) is highly stable, whereas the first overtone (orange circles) exhibits erratic fluctuations characteristic of highly transient, un-resonated local excitations.}
         \label{fig:deviation-im}
     \end{subfigure}
    \caption{MPM spectral extraction of the first echo as a function of cavity length $L \in [110, 210]$. Frequency shifts $\Delta \omega$ are measured relative to the fundamental Schwarzschild QNM.}
    \label{fig:deviation}
\end{figure}

Consistent with this severe spectral aliasing, 
full-wave time domain numerical extractions unequivocally reject the aforementioned steady state blue-shift prediction.
As shown in Fig.\ \ref{fig:deviation-re}, the effective central frequency extracted via the Matrix Pencil Method (MPM) exhibits absolutely no $1/L$ blue-shift. 
Instead, it remains rigidly locked near the Schwarzschild fundamental ringdown frequency ($\Delta\omega_c \simeq 0$). 
This phenomenologically verifies that the first echo is utterly unbound by the global pole quantization; 
the non-zero phase shift $\delta_{\mathrm{tot}}$ is instead fully converted into the dispersive arrival time gliding via the Wigner delay relation \cite{Wigner:1955zz}, exactly as analytically predicted in Sec.\ \ref{sec:IV}.

Furthermore, Fig.\ \ref{fig:deviation-im} reveals that while the macroscopic fundamental mode remains stable, the imaginary part of the first overtone (orange circles) exhibits severe, length-dependent fluctuations. 
Such erratic behavior is fundamentally incompatible with an established steady state cavity mode, proving that these overtones are highly transient excitations governed by local, short-lived coherence conditions. 
The large positive shift in their imaginary parts confirms that these high-order components dissipate overwhelmingly before any global resonance can be established.
This provides independent methodological proof that early low-finesse echoes are purely transient scattering packets.

\begin{figure}[!htb]
    \centering
    \begin{subfigure}[t]{0.45\textwidth}
         \centering
         \includegraphics[width=\textwidth]{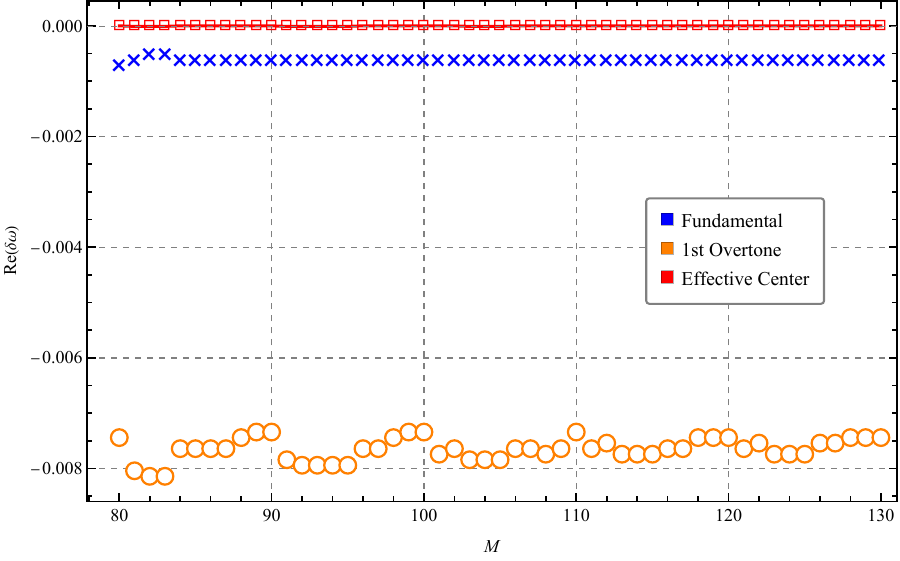}
         \caption{Real part deviation $\Re(\delta\omega)$. The fundamental mode (blue) precisely matches the Schwarzschild baseline, while the overtone (orange) displays a stable, barrier-induced negative shift.}
         \label{fig:stability-re}
     \end{subfigure}
     \begin{subfigure}[t]{0.45\textwidth}
         \centering
         \includegraphics[width=\textwidth]{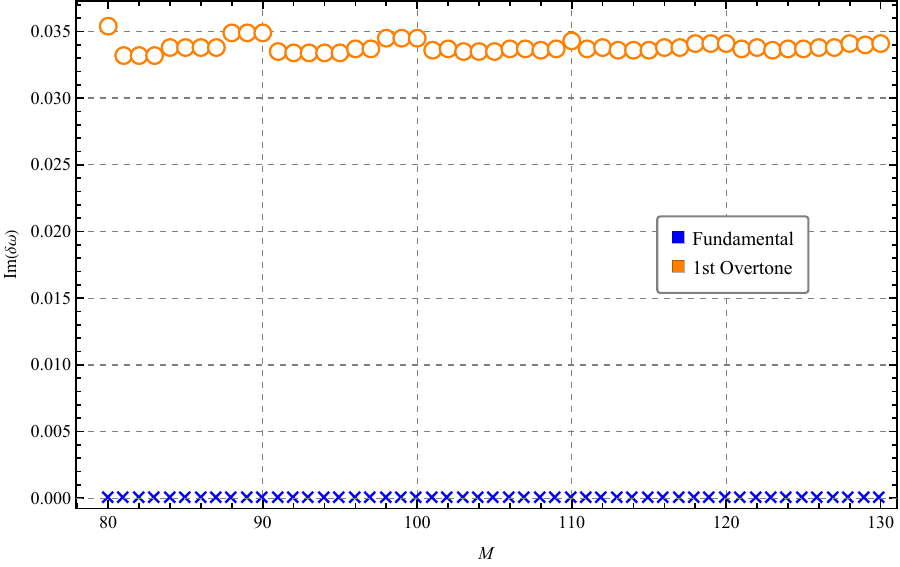}
         \caption{Imaginary part deviation $\Im(\delta\omega)$. The stable positive shift confirms the physical presence of the rapidly dissipating overtone component.}
         \label{fig:stability-im}
     \end{subfigure}
    \caption{Numerical convergence test of the MPM spectral extraction with respect to the matrix pencil order $M \in [80, 130]$ for the $L=180$ benchmark sample.}
    \label{fig:stability}
\end{figure}

To rigorously preclude numerical artifacts, we perform a convergence test on the MPM order parameter $M \in [80, 130]$ for the $L=180$ benchmark model. 
Fig.\ \ref{fig:stability} demonstrates exceptional numerical stability across all extracted modes.

The real part deviation of the fundamental mode (Fig.\ \ref{fig:stability-re}, blue crosses) is negligible ($\sim 10^{-4}$), re-confirming locking to the unperturbed Schwarzschild geometry.
The first overtone (orange circles) exhibits a highly stable non-zero shift ($\Re(\Delta\omega) \sim -0.008$), proving its existence is physically real and not an algorithmic artifact, likely reflecting a subtle barrier-induced effective potential correction. 
Similarly, its imaginary part (Fig.\ \ref{fig:stability-im}) stabilizes at a positive deviation ($\Im(\Delta\omega) \simeq 0.035$). 
This consistent convergence definitively proves that the highly dissipative, un-resonated overtone is an intrinsic physical component of the transient scattering packet, strictly validating the physical conclusions derived from Fig.\ \ref{fig:deviation}.

\section{Physical distinction between the cavity-formation threshold and the time domain observability threshold}\label{app:B}

In Sec.\ \ref{sec.III} of the main text, based on the direct comparison between the resonance linewidth and the free spectral range, we derived the effective reflectivity threshold for the occurrence of energy level aliasing in the comb spectrum, i.e., $R_{A,\mathrm{eff}} \lesssim \me^{-\pp}$ or $\me^{-2\pp}$.
However, when we discuss the observability of cavity modes in waveform from the perspective of cavity lifetime, we obtain another threshold $R_{A,\mathrm{eff}} \sim \mathcal{O}(10^{-1})$.
This appendix aims to clearly distinguish the different physical benchmarks corresponding to these two thresholds, prove that they are not physically contradictory.

We investigate the behavior of a wave packet traveling between two potential barriers separated by $L$.
The effective reflectivity for a single round trip of the wave packet is still denoted as $R_{A,\mathrm{eff}}$, and the characteristic round-trip period is roughly $\bar{T}_{\mathrm{round}} \simeq 2L$.
In the geometric optics limit, the amplitude decay of the wave packet after $n$ round trips is proportional to $R_{A,\mathrm{eff}}^n$. 
Mapping it to the evolution function at continuous time $t \simeq n \bar{T}_{\mathrm{round}}$, we obtain the exponential decay law of the wave packet amplitude
\begin{equation}
    |\Psi(t)| \propto R^{\frac{t}{\bar{T}_{\mathrm{round}}}} \equiv \exp\left( -\frac{t}{\tau} \right),
\end{equation}
where we define the characteristic photon lifetime of the system
\begin{equation}\label{eq:lifetime}
    \tau = \frac{\bar{T}_{\mathrm{round}}}{|\log R_{A,\mathrm{eff}}|}.
\end{equation}
Physically, for a stable coherent standing wave to be established in the resonant cavity, the trapped energy must survive long enough to complete at least a full physical round trips.
This condition requires that the lifetime of the wave packet should be approximately on the order of the round-trip period, i.e., $\tau \gtrsim \xi \bar{T}_{\mathrm{round}}$, where $\xi \sim \mathcal{O}(1)$ is an empirical constant measuring the number of round trips required to maintain an effective closed loop.
Substituting into the lifetime expression \eqref{eq:lifetime}, we obtain an order-of-magnitude estimate for the survival threshold
\begin{equation}
    R_{A,\mathrm{eff}} \geq \me^{-1/\xi}.
\end{equation}
Taking the loosest limit $\xi=1$, i.e., the wave packet can only complete one round trip, we have $R_{A,\mathrm{eff}} \gtrsim \me^{-1} \simeq 0.37$.
This indicates that for the resonant cavity mode to be observable in the time domain waveform, the effective reflectivity of the system must reach at least the order of $\mathcal{O}(10^{-1})$.

These two thresholds govern fundamentally distinct physical conditions:
\begin{enumerate}
    \item {\em Time domain Survival Threshold} ($\mathcal{O}(10^{-1})$):
    Dictates whether the wave packet retains sufficient energy to complete a single effective coherent cycle. 
    Below this limit, the system undergoes catastrophic single-trip leakage, completely precluding the establishment of long-lived interference patterns in full waveform.
    \item {\em Frequency domain Resolvability Threshold} ($\mathcal{O}(10^{-2})$):
    Dictates spectral aliasing. 
    Even if mathematical poles exist, a reflectivity below this bound guarantees that their corresponding Lorentzian linewidths overlap, merging into an unresolvable continuum.
\end{enumerate}

To explicitly demonstrate genuine steady state behavior, we construct a high-finesse square double-barrier toy model ($L=180$, height $V=10$, width $w=3$), achieving near-total reflection ($|R_{A,{\rm eff}}| \simeq 0.99 \sim \mathcal{O}(1)$).

\begin{figure}[!htb]
    \centering
    \begin{subfigure}[t]{0.45\textwidth}
         \centering
         \includegraphics[width=\textwidth]{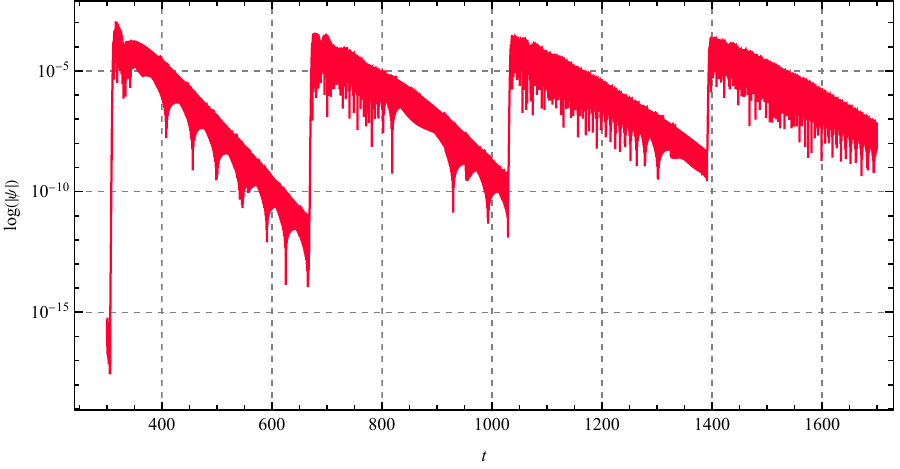}
         \caption{Time domain waveform of the double-barrier toy model ($|R_{A,{\rm eff}}| \simeq 0.99$). The signal exhibits sustained standing-wave ringing with negligible decay over thousands of $M$, satisfying the survival threshold.}
         \label{fig:waveform-barrier}
     \end{subfigure}
     \begin{subfigure}[t]{0.45\textwidth}
         \centering
         \includegraphics[width=\textwidth]{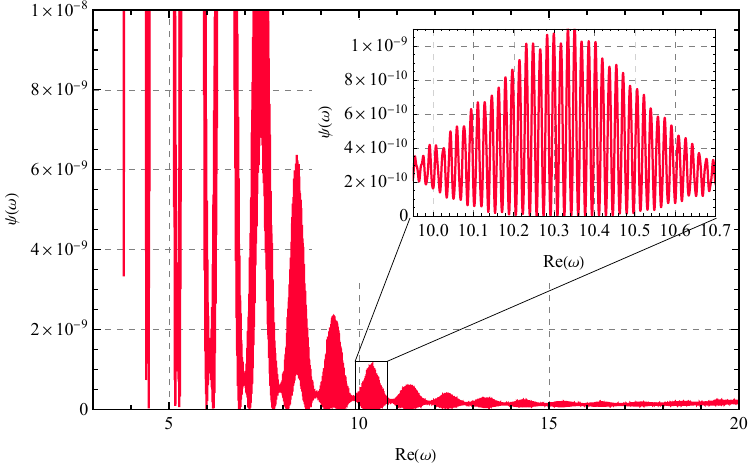}
         \caption{Global Fourier spectrum displaying extremely sharp, well-separated peaks. The inset highlights the fine structure of the Lorentzian combs, whose full-width at half-maximum is vastly narrower than the free spectral range.}
         \label{fig:FFT-barrier}
     \end{subfigure}
    \caption{The hallmark of genuine steady state resonance in the high-finesse regime ($L=180, V=10, w=3$).}
    \label{fig:barrier}
\end{figure}

As shown in Fig.\ \ref{fig:barrier}, this high-finesse regime naturally produces sustained, slowly decaying time domain standing waves and a highly resolved frequency domain Lorentzian comb. 
This control model definitively verifies that the absence of comb spectra in our realistic astrophysical models (Sec.\ \ref{sec:II}) is an intrinsic physical consequence of extreme low-finesse dissipation.

\section{Next-to-leading order corrections to the dynamic dispersive analytic kernel}\label{app:C}

In Sec.\ \ref{sec:IV}, our leading-order dispersive theory truncates the logarithmic amplitude $P_n(\omega)$ at second order, while extending the phase $\Theta_n(\omega)$ to third order to capture the asymmetric tails. 
Any mismatch between the $\omega_{c,n}$ and the $\omega_{{\rm peak},n}$ enters only beyond leading order, and is encoded by the Edgeworth-type higher-order amplitude corrections discussed below.
Retaining the third-order amplitude expansion within the exponential kernel to capture spectral skewness is mathematically prohibited—any odd-order polynomial in the real exponent irreparably destroys the absolute integrability of the inverse Fourier transform over the real frequency axis.
This appendix establishes the rigorous next-to-leading-order framework. 
We demonstrate that amplitude skewness must instead be factored out of the exponent, natively manifesting in the time domain as an Edgeworth-type perturbative correction to the leading-order wave packet.

According to the definition in the main text, the time domain signal of the $n$-th echo is given by the inverse Fourier transform \eqref{eq:fourier-inverse}.
We expand the logarithmic amplitude $P_n(\omega)$ to higher orders in the neighborhood of the central frequency $\omega_{c,n}$,
\begin{equation}
    P_n(\omega) \simeq P_n(\omega_{c,n}) - \frac{\Delta\omega^2}{2\sigma_{\omega,n}^2} + \frac{\kappa_{3,n}}{6}\Delta\omega^3 + \frac{\kappa_{4,n}}{24}\Delta\omega^4 + \cdots,
\end{equation}
where the higher-order expansion coefficients are defined as $\kappa_{m,n} \equiv P_n^{(m)}(\omega_{c,n})$.
Suppose we keep the third-order term $\kappa_{3,n}\Delta\omega^3/6$ directly inside the exponential function just like the phase term, then the integral kernel that determines the convergence of the inverse Fourier transform becomes
\begin{equation}
    \exp\left[ P_n(\omega_{c,n}) - \frac{\Delta\omega^2}{2\sigma_{\omega,n}^2} + \frac{\kappa_{3,n}}{6}\Delta\omega^3 \right].
\end{equation}
Since $\Delta\omega^3$ is an odd function, its signs on the two sides of the real frequency axis are necessarily opposite.
In the limit $|\Delta\omega| \to \infty$, the absolute growth rate of the cubic term will inevitably overwhelm the suppressing quadratic term.
Therefore, as long as $\kappa_{3,n} \neq 0$, there must exist on one side of the real axis
\begin{equation}
    -\frac{\Delta\omega^2}{2\sigma_{\omega,n}^2} + \frac{\kappa_{3,n}}{6}\Delta \omega^3 \to \infty.
\end{equation}
This indicates that the inverse Fourier transform integral containing odd higher-order amplitude terms loses absolute integrability.

Attempting to regularize this divergence via ad hoc finite frequency cutoffs is theoretically invalid. 
Such procedures unlawfully extrapolate a local Taylor expansion across the entire real axis, inevitably contaminating the time domain reconstruction with severe boundary artifacts. 
Conversely, the cubic phase term $\exp(\mi K_n \Delta\omega^3 / 6)$ remains strictly unitary over the full frequency domain. 
It purely governs the interference dispersion to generate the asymmetric tails, without compromising the fundamental Gaussian convergence kernel. 
This profound mathematical asymmetry between amplitude and phase strictly mandates the aforementioned Edgeworth perturbative framework.

Since the higher-order amplitude terms describing the skewness of the wave packet's frequency domain distribution cannot be placed inside the exponential, the only rigorous mathematical approach to retain this physical information is to treat the skewness as a perturbation controlled by a small parameter and perform a polynomial expansion outside the dominant Gaussian kernel.
We define the higher-order remainder of the amplitude expansion as
\begin{equation}
    R_n(\Delta\omega) \equiv \frac{\kappa_{3,n}}{6}\Delta\omega^3 + \frac{\kappa_{4,n}}{24}\Delta\omega^4 + \cdots.
\end{equation}
The frequency domain amplitude factor can be decomposed as
\begin{equation}
    \exp \left( P_n(\omega) \right) = \mathcal{A}_n \exp\!\left( -\frac{\Delta\omega^2}{2\sigma_{\omega,n}^2} \right) \exp \left( R_n(\Delta\omega) \right).
\end{equation}
To define the asymptotic validity of this decomposition, we introduce the dimensionless frequency shift and define the dimensionless higher-order skewness coefficients
\begin{equation}
    \alpha_{3,n} \equiv \frac{\kappa_{3,n}\sigma_{\omega,n}^3}{6}, \qquad \alpha_{4,n} \equiv \frac{\kappa_{4,n}\sigma_{\omega,n}^4}{24}.
\end{equation}
Taking $y \equiv \Delta\omega / \sigma_{\omega,n}$, as long as the higher-order expansion coefficients satisfy the perturbative condition $|\kappa_{m,n}/\sigma_{\omega,n}^m| \ll 1$ for any $m$ within the dominant energy band $|y| \sim \mathcal{O}(1)$, we can expand the higher-order exponential term in a series
\begin{equation}
    \exp \left( R_n(\Delta\omega) \right) \simeq 1 + \frac{\kappa_{3,n}}{6}\Delta\omega^3 + \frac{\kappa_{4,n}}{24}\Delta\omega^4 + \cdots.
\end{equation}
Substituting this polynomial product back into the original integrand, we obtain the next-to-leading-order frequency domain structure: a Gaussian main kernel that guarantees absolute integrability, superimposed with a series of higher-order polynomial correction factors.

Finally, we transform the above polynomial correction in the frequency domain to the time domain to intuitively clarify its specific correction effect on the waveform.
Let $\Psi_n^{\mathrm{LO}}(t)$ denote the convolution of the complex Gaussian kernel and the Airy kernel derived in Sec.\ \ref{sec:IV} of the main text.
Define the local time coordinate $x \equiv t - t_n$, and using the derivative theorem of Fourier transform, the full-order echo wave packet including higher-order amplitude skewness correction can be expressed in the time domain as
\begin{equation}
    \Psi_n^{\mathrm{NLO}}(t) \simeq \left[ 1 + \frac{\kappa_{3,n}}{6}(\mi \partial_x)^3 + \frac{\kappa_{4,n}}{24}(\mi \partial_x)^4 + \cdots \right] \Psi_n^{\mathrm{LO}}(t).
\end{equation}
To reveal the physical meaning of the action of this differential operator on the integral kernel, we temporarily neglect the third-order dispersion ($K_n \to 0$), i.e., take the integral kernel as a pure complex Gaussian kernel.
At this point, the leading-order integral kernel reduces to the chirped Gaussian form $\Psi_n^{\mathrm{LO}}(t) \propto \exp(-x^2 / 2Q_n)$, where $Q_n = \sigma_{\omega,n}^{-2} - \mi D_n$.
Define the complex dimensionless variable $z \equiv x / \sqrt{Q_n}$, and using the probabilistic expression of the Hermite polynomials, we obtain the corrected expression of the time domain wave packet
\begin{equation}
    \Psi_n^{\mathrm{NLO}}(t) \simeq
    \Psi_n^{\mathrm{LO}}(t) \left[ 1 + \mi \frac{\kappa_{3,n}}{6 Q_n^{3/2}} \mathrm{He}_3(z) + \frac{\kappa_{4,n}}{24 Q_n^2} \mathrm{He}_4(z) + \cdots \right],
\end{equation}
where $\mathrm{He}_n(x)$ is the $n$-th order Hermite polynomial.

In the case of the complete integral kernel given in the main text, the above correction method still holds, but its analytical expression can no longer be written using special functions.
The physical effect of the correction terms on the time domain wave packet is that the even-order derivative operators fine-tune the effective width and kurtosis of the main peak, while the odd-order derivative operators introduce local left-right asymmetry and adjust the bias and tails of the waveform.

\section*{Acknowledgement}

H.-W.Hu is grateful for C.Lan for useful discussions. 
This work is supported by the National Natural Science Foundation of China No.~12475067 and No.~12235019.

\bibliographystyle{utphys}
\bibliography{citation}

\end{document}